\documentclass[a4,amssymb,amsmath,aps,prb,reprint,superscriptaddress,showpacs,floatfix]{revtex4-1}
\usepackage{amssymb,amsmath}
\usepackage{color}
\usepackage{graphicx}
\usepackage{epstopdf}
\usepackage{array}
\usepackage{braket}
\usepackage{tikz}
\usepackage{pgfplots}
\usepackage{bm}
\usepackage{hyperref}
\allowdisplaybreaks
\hypersetup{colorlinks=true, citecolor=blue, urlcolor=blue, linkcolor=red}

\newcommand{\naoso}{NaOsO$_{\text{3}}$}
\newcommand{\cdoso}{Cd$_{\text{2}}$Os$_{\text{2}}$O$_{\text{7}}$}

\begin{document}
\title{Crossover from itinerant to localized magnetic excitations through the metal-insulator transition in \naoso}

\author{J. G. Vale}
\email{j.vale@ucl.ac.uk}
\affiliation{London Centre for Nanotechnology and Department of Physics and Astronomy, University College London, Gower Street, London, WC1E 6BT, United Kingdom}
\affiliation{Laboratory for Quantum Magnetism, \'{E}cole Polytechnique F\'ed\'erale de Lausanne (EPFL), CH-1015, Switzerland}

\author{S. Calder}
\email{caldersa@ornl.gov}
\affiliation{Neutron Scattering Division, Oak Ridge National Laboratory, Oak Ridge, Tennessee 37831, USA}

\author{C. Donnerer}
\affiliation{London Centre for Nanotechnology and Department of Physics and Astronomy, University College London, Gower Street, London, WC1E 6BT, United Kingdom}

\author{D. Pincini}
\affiliation{London Centre for Nanotechnology and Department of Physics and Astronomy, University College London, Gower Street, London, WC1E 6BT, United Kingdom}
\affiliation{Diamond Light Source, Harwell Science and Innovation Campus, Didcot, Oxfordshire, OX11 0DE, United Kingdom}

\author{Y. G. Shi}
\affiliation{Beijing National Laboratory for Condensed Matter Physics and Institute of Physics, Chinese Academy of Sciences, Beijing 100190, China}
\affiliation{Research Center for Functional Materials, National Institute for Materials Science, 1-1 Namiki, Tsukuba, Ibaraki 305-0044, Japan}

\author{Y. Tsujimoto}
\affiliation{Research Center for Functional Materials, National Institute for Materials Science, 1-1 Namiki, Tsukuba, Ibaraki 305-0044, Japan}

\author{K. Yamaura}
\affiliation{Research Center for Functional Materials, National Institute for Materials Science, 1-1 Namiki, Tsukuba, Ibaraki 305-0044, Japan}
\affiliation{Graduate School of Chemical Sciences and Engineering, Hokkaido University, North 10 West 8, Kita-ku, Sapporo, Hokkaido 060-0810, Japan}

\author{M. Moretti Sala}
\affiliation{ESRF, The European Synchrotron, 71 Avenue des Martyrs, 38043 Grenoble, France}

\author{J. van den Brink}
\affiliation{Institute for Theoretical Solid State Physics, IFW Dresden, D01171 Dresden, Germany}

\author{A. D. Christianson}
\affiliation{Neutron Scattering Division, Oak Ridge National Laboratory, Oak Ridge, Tennessee 37831, USA}
\affiliation{Department of Physics and Astronomy, University of Tennessee, Knoxville, TN 37996, USA}

\author{D. F. McMorrow}
\affiliation{London Centre for Nanotechnology and Department of Physics and Astronomy, University College London, Gower Street, London, WC1E 6BT, United Kingdom}


\begin{abstract}
{\naoso\ undergoes a metal-insulator transition (MIT) at 410~K, concomitant with the onset of antiferromagnetic order. The excitation spectra have been investigated through the MIT by resonant inelastic x-ray scattering (RIXS) at the Os L$_{\text{3}}$ edge. Low resolution ($\Delta E \sim \text{300~meV}$) measurements over a wide range of energies reveal that local electronic excitations do not change appreciably through the MIT. This is consistent with a picture in which structural distortions do not drive the MIT.
In contrast, high resolution ($\Delta E \sim \text{56~meV}$) measurements show that the well-defined, low energy magnons in the insulating state weaken and dampen upon approaching the metallic state. Concomitantly, a broad continuum of excitations develops which is well described by the magnetic fluctuations of a nearly antiferromagnetic Fermi liquid.  By revealing the continuous evolution of the magnetic quasiparticle spectrum as it changes its character from itinerant to localized, our results provide unprecedented insight into the nature of the MIT in \naoso. In particular, the presence of weak correlations in the paramagnetic phase implies a degree of departure from the ideal Slater limit.\footnote{This work was prepared as a joint submission with Physical Review Letters; please see Ref.~\onlinecite{vale2018_prl} and references therein.}}
\end{abstract}
\maketitle

Competing interactions are a fundamental driving force governing the electronic and magnetic properties of transition metal oxides (TMOs). The metal-insulator transition (MIT) is a prime example of a phenomenon which is driven by this competition, and consequently remains a source of significant experimental and theoretical interest.\cite{mott1990, edwards1995, imada1998, cooper2003, dobrosavljevic2011}
In 3d TMOs, the presence of a MIT is governed primarily by an effective interaction strength $U/t$, where $U$ represents inter-electronic Coulomb repulsion, and $t$ is inter-site hopping. This corresponds to the well-studied Mott-Hubbard paradigm, where MITs can be driven by bandwidth or band-filling control. 

Yet in 5d TMOs, the more delocalized valence orbitals, along with strong spin-orbit coupling (SOC), gives rise to new phenomenology. For example, SOC in $5d^5$ iridates affects the ground state to such an extent that even a moderate $U$ is sufficient to open up a Mott-like insulating gap.\cite{witczak-krempa2014, rau2016}
Furthermore, a number of MITs have been observed which are intimately entwined with the onset of long-ranged, commensurate antiferromagnetic order. Notably these MITs do not appear to be associated with any spontaneous structural symmetry breaking, placing them outside of the aforementioned Mott-Hubbard paradigm. Examples include some of the 5$d^5$ pyrochlore iridates  R$_{\text{2}}$Ir$_{\text{2}}$O$_{\text{7}}$ (R = Ln$^{\text{3+}}$),\cite{matsuhira2011, nakayama2016} plus the 5$d^3$ osmates \cdoso,\cite{mandrus2001, padilla2002, yamaura2012, hiroi2015} and \naoso. Various mechanisms have been proposed to describe these MITs, notably the Lifshitz and Slater mechanisms.

\begin{figure}[t]
\centering
\includegraphics{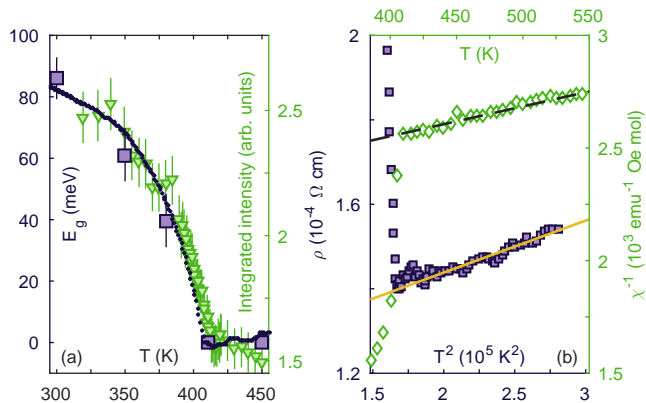}
\caption{Summary of magnetic and electronic behavior in \naoso\ as a function of temperature. (a): Charge gap extracted from resistivity data (dotted line),\cite{shi2009} optical gap (purple squares),\cite{lovecchio2013}, and integrated intensity of $(0,\,1,\,1)$ magnetic Bragg peak obtained from powder neutron diffraction (green triangles).\cite{calder2012} The MIT and onset of antiferromagnetic order appear to be intimately linked.
(b): Resistivity $\rho$ versus $T^2$ (purple squares), and inverse susceptibility $\chi^{-1}$ versus $T$ (green diamonds). Overlaid are fits to the theoretical expressions for the resistivity of a Fermi liquid (solid line), and a Curie-Weiss paramagnet (dashed line). Data and fitted parameters taken from Ref.~\onlinecite{shi2009}. }
\label{summary}
\end{figure}

A Lifshitz transition involves a change of topology of the Fermi surface at $T_{\text{MI}}$, and is an example of a quantum phase transition (QPT) at $T=0$. At finite temperatures, the theoretically expected singularities in thermodynamic parameters -- such as the thermal expansion coefficient -- become washed out, and the QPT is in fact a crossover between the `ordered' and `disordered' phases. 
Meanwhile in a Slater insulator, the onset of antiferromagnetic order itself drives the onset of an insulating gap below the N\'{e}el temperature. In the most general sense, insulating behavior arises from an ordered magnetic exchange field governed by mean-field type interactions. True Slater insulating behavior -- as defined in the original theoretical works -- is limited to systems with a half-filled $t_{2g}$ manifold.\cite{slater1951, descloizeaux1959} 
A number of works have proposed that \naoso\ is a rare example of a Slater insulator in three dimensions (Fig.~\ref{summary}a).\cite{shi2009, calder2012, du2012, jung2013, lovecchio2013} 
Together with significant spin-phonon coupling,\cite{calder2015_naoso3} one observes an unprecedented connection between the magnetic, electronic, structural, and phonon degrees of freedom in \naoso. 

There are, however, a number of outstanding questions with regards to the true nature of the MIT in \naoso.
Optical conductivity measurements reveal a continuous opening of the electronic gap with decreasing temperature $\left\lbrace\Delta_g (0) = \text{102(3)~meV}\right\rbrace$, and an MIT in which electronic correlations play a limited role. This is consistent with a Slater picture in which interactions are mean-field like.\cite{lovecchio2013}
Meanwhile previous RIXS measurements showed well-defined and strongly gapped ($\sim\!\text{50~meV}$) dispersive spin-wave excitations at 300~K.\cite{calder2017_naoso3}  This was found to be consistent with an anisotropic nearest-neighbour Heisenberg picture for the magnetic Hamiltonian, and is suggestive of localized magnetic moments. 
Furthermore recent density functional theory (DFT) calculations \cite{bongjaekim2016} suggest that the Fermi surface may be reconstructed at the MIT by magnetic fluctuations of itinerant Os moments; in a so-called spin-driven Lifshitz MIT.
The question remains whether either the spin or electronic excitations remain coherent through the MIT, and if there is any evidence of coupling to any of the other relevant degrees of freedom present in the system. 

In this manuscript, we establish that there is a continuous progression from itinerant to localized behavior through the MIT in \naoso. This is revealed by a significant renormalization of the magnetic quasi-particle spectral weight over large ranges of momentum and energy transfer.
In particular, the presence of correlations in the metallic state immediately leads to a deviation from mean-field behavior, and hence, true Slater phenomenology. 

Our experiments relied on exploiting the unique ability of RIXS to provide momentum-resolved 
sensitivity to the excitations of the orbital, electronic and magnetic degrees of freedom.
By providing data on an experimental test case, in which an effective interaction strength can be tuned simply by varying the temperature, our work in turn helps extend the utility of RIXS, which has hitherto been best understood in the localized limit.\cite{haverkort2010, jia2014, kim_khaliullin2017}

\section{Experimental setup and sample geometry}\label{setup_geometry}
RIXS measurements were performed at the Os $L_3$ edge ($E = \text{10.871~keV}$) on the ID20 spectrometer at the ESRF, Grenoble.\cite{moretti2018} Preliminary measurements were performed at 9-ID-B, Advanced Photon Source.
For high resolution measurements a Si $(\text{6},\,\text{6},\,\text{4})$ channel-cut secondary monochromator was used to select the incident energy. A Si $(\text{6},\,\text{6},\,\text{4})$ diced spherical analyser (2~m Rowland circle radius, 60~mm diameter) was used to reflect the scattered photons towards a Maxipix CCD detector (pixel size 55~$\mu$m) and discriminate the scattered photon energy. The total energy resolution was determined to be $\Delta$E = 56~meV, based on diffuse scattering from a polypropylene-based adhesive tape (Fig.~\ref{res_ftn}).
For low resolution measurements, a similar setup was used, but with a Si $(\text{3},\,\text{1},\,\text{1})$ channel-cut secondary monochromator instead. The total energy resolution in this case was $\Delta$E $\approx \text{300~meV}$.

\begin{figure}[t]
\includegraphics{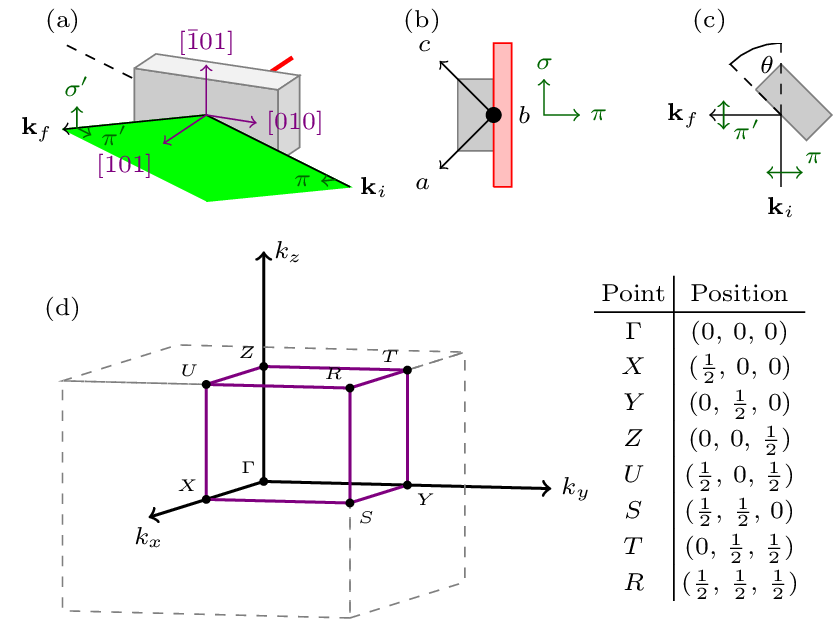}
\caption{(a): Schematic of experimental geometry. Magnetic moments lie along the $c$-axis, the $b$-axis lies approximately within the scattering plane. (b): View for $\theta=0$ parallel to the direction of the incident beam. (c): Top-down view. In the experimental geometry $\theta\approx 60^{\circ}$. (d): Conventional orthorhombic unit cell for the $Pnma$ space group (gray dashed) overlaid with the Brillouin zone (solid purple). Highlighted are the high symmetry points of the Brillouin zone, which are listed in the neighbouring table.}
\label{geometry}
\end{figure}

A single crystal of \naoso\ (approximate dimensions 0.3$\times$0.3$\times$0.3~mm$^{\text{3}}$) was oriented such that the $(\text{1},\,\text{0},\,\text{1})$ direction was normal to the sample surface (Fig.~\ref{geometry}). The sample was mounted with silver paint and placed in a custom-made heater setup filled with helium exchange gas. Temperature stability was better than $\pm \text{0.5}$~K.
The scattering plane and incident photon polarization were both horizontal ($\pi$--incident polarization) with the incident beam focussed to a size of 20$\times$10~$\mu$m$^{\text{2}}$ (H$\times$V) at the sample position.

\section{Temperature dependence of orbital excitations}\label{orbital_excitations}

\begin{figure*}
\includegraphics{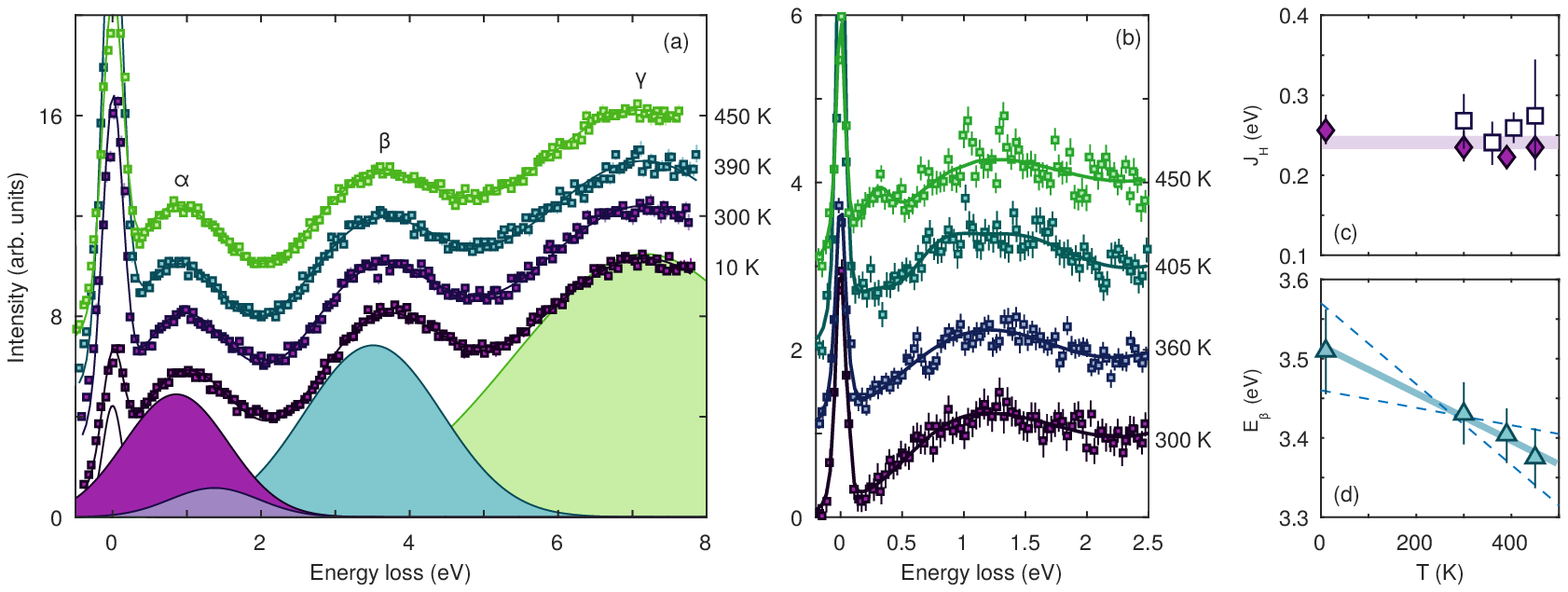}
\caption{(a): Orbital excitations as a function of temperature, obtained using a low resolution setup $\left(\Delta E = \text{300~meV}\right)$. The excitations are practically temperature independent. Solid lines are best fit to $LS$-coupling model described in the main text. Shaded peaks show the relative contributions at 10~K. 
(b): High-resolution RIXS spectra ($\Delta E = \text{56~meV}$) obtained as a function of temperature close to $\Gamma$ $(\text{5.05},\,\text{3.05},\,\text{4.05})$. Solid line is best fit to $LS$-coupling model.
(c): Value of Hund's coupling $J_{\text{H}}$ calculated from fitted peak positions of intra-$t_{2g}$ excitations at $3J_H$ and $5J_H$ (peak A). Filled (open) symbols: low (high) resolution data. Solid line shows best fit to combined data, with $J_H = \text{0.25(1)~eV}$. (d): Fitted energy of peak $\beta$ as function of temperature. Solid line is linear fit to the data. Dashed lines reflect the significant uncertainty in this parameter.
}
\label{comp_full}
\end{figure*}

We start with the orbital excitations as a function of temperature. A low resolution setup $\left(\Delta E = \text{300~meV}\right)$ was used in order to examine the orbital excitations out to large energy loss ($\sim$ 8~eV). Selected spectra are displayed in Fig.~\ref{comp_full}a.
Just as in Ref.~\onlinecite{calder2017_naoso3}, four peaks are evident in the RIXS spectra. The peak centered at zero energy loss comprises the elastic line and other low energy features such as phonons, magnons etc. The three remaining peaks refer to excitations either: within the $t_{2g}$ manifold ($\bm{\alpha}$), from $t_{2g}^3\rightarrow t_{2g}^2e_g$ states ($\bm{\beta}$), or $t_{2g}^3 \rightarrow t_{2g}e_g^2$ and ligand-to-metal charge transfer excitations ($\bm{\gamma}$). This assignment follows the previous RIXS measurements on \naoso\ and \cdoso,\cite{calder2016_cdoso, calder2017_naoso3} as well as quantum chemistry calculations.\cite{bogdanov2013}

At first inspection, the intensity and positions of peaks $\bm{\alpha}$--$\bm{\gamma}$ appear to be essentially temperature independent. Moreover, the individual features are much broader than the instrumental resolution, indicative of a system with significant non-local character. 
High resolution $(\Delta E = \text{55~meV})$ measurements over a limited range show similar behavior (Fig.~\ref{comp_full}b); peak $\bm{\alpha}$ remains broad and relatively featureless. A degree of fine structure can be ascertained around 1~eV -- especially at higher temperature -- but unfortunately the statistical quality of the data is insufficient to draw firm conclusions. The noticeable low energy ($<\text{0.5~eV}$) behavior will be addressed later in this manuscript.
Contrast the RIXS spectra presented here with the isoelectronic Ca$_3$LiOsO$_6$ and Ba$_2$YOsO$_6$.\cite{taylor2017} In these materials, OsO$_{6}$ octahedra are separated from each other by Li (Y). The excitations are thus more localized in real space, with a correspondingly longer lifetime and narrower bandwidth.

\subsection*{Calculation of microscopic parameters}
It is possible to go further and obtain quantitative conclusions from the data using a ligand field model.
The magnitudes of Hund's coupling $J_{\text{H}}$ and the one-electron spin-orbit coupling parameter $\zeta$ can be estimated from the energies of the electronic transitions in the RIXS spectra. The starting point is a single Os$^{5+}$ cation octahedrally coordinated to six O$^{2-}$ ions, assuming no distortion away from ideal octahedral symmetry. For \naoso\ this is not strictly true, however the distortion is sufficiently small ($D\tau/Dq\approx -0.01$, $Ds,Dt\rightarrow 0$ within uncertainty) that this approximation holds within experimental resolution.\cite{hempel1976, calder2012}
Assuming a spherically symmetric Coulomb interaction, and $t_{2g}$ electronic wavefunctions obtained from crystal field theory which have pure $d$-character (i.e.~no hybridisation), then the Hund's coupling $J_{\text{H}}$ can be expressed in terms of the Racah parameters: $J_{\text{H}}=3B+C$.\footnote{The Racah parameters can be directly related to Slater integrals using standard transformations: $A = F_0 - 49F_4$, $B = F_2 - 5F_4$, $C = 35F_4$. Note that $F_k$ and $F^k$ -- used at various points within the literature --  are distinct quantities: $F_2 = F^2/49$, $F_4=F^4/441$. Strictly speaking, this expression is only correct for coupling within the $t_{2g}$ manifold; the Hund's coupling for electrons between $t_{2g}$ and $e_g$ orbitals, or within the $e_g$ manifold, is given by $J_{\text{H}}=4B+C$. \cite{sugano1970} Our estimates are based solely on peak $\bm{\alpha}$, that is, purely within the $t_{2g}$ manifold, so $J_{\text{H}}=3B+C$ holds.}

Within a simple $LS$-coupling model, which neglects the effects of weak octahedral distortion, peak $\bm{\alpha}$ is comprised of two overlapping excitations centred at 3$J_{\text{H}}$ and 5$J_{\text{H}}$ (Fig.~\ref{multiplet_levels}a). Hence we attempted to quantify the microscopic parameters to first order by fitting the low resolution ($\Delta E = \text{300~meV}$) RIXS data to a sum of five Gaussians.
We find that $J_{\text{H}}=\text{0.24(1)~eV}$ is temperature independent within experimental uncertainty (Fig.~\ref{comp_full}c). 
Fitting high resolution $(\Delta E = \text{55~meV})$ data with the same model gives a similar result: $J_H = \text{0.26(2)~eV}$.
Both values for $J_H$ compare well with estimates obtained from other $5d^3$ materials.\cite{calder2016_cdoso, taylor2017}

\subsection*{Discussion}
The assumption that the orbital excitations can be described in terms of four Gaussians is incomplete.
Assuming ideal octahedral symmetry, $LS$-coupling, and zero spin-orbit coupling, then peak $\bm{\alpha}$ is indeed comprised of two overlapping excitations at $3J_{H}$ and $5J_{H}$. However peak $\bm{\beta}$ is in fact made up of ten overlapping multiplets (${}^2A_1$, ${}^2A_2$, $2{}^2T_1$, $2{}^2T_2$, $2{}^2E$, ${}^4T_1$, ${}^4T_2$), with their relative energies dictated primarily by $B$ and $C$.\cite{sugano1970} 
\begin{figure}[t]
\centering
\includegraphics{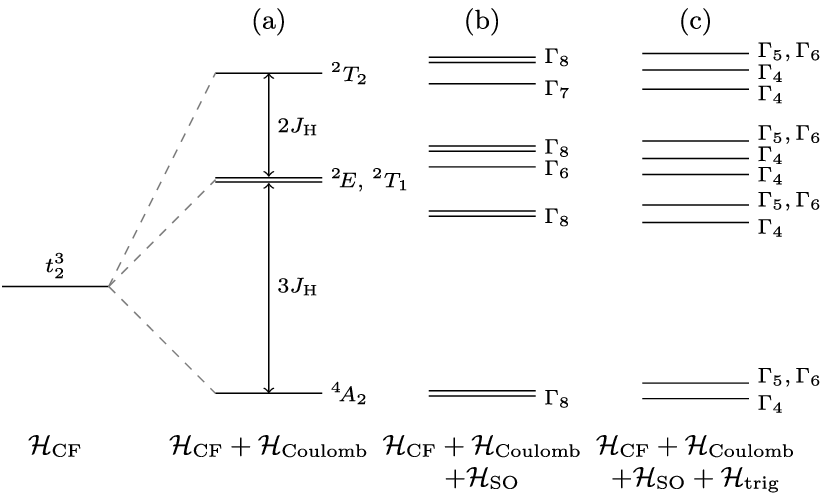}
\caption{(a): Terms of a $t_2^{\text{3}}$ configuration in a first (strong field) approximation, including the effect of Coulomb repulsion. Spin-orbit coupling and distortions away from an ideal octahedral geometry are neglected. The Hund's coupling $J_{\text{H}}$ is equivalent to $\text{3}B+C$, where $B$ and $C$ are Racah parameters. The terms ${}^2\!E$ and ${}^2T_1$ are accidentally degenerate. (b): Electronic fine structure levels obtained when spin-orbit coupling is added as a weak perturbation. Levels are calculated by decomposition of the direct product $\Gamma \times D^{(\text{3/2})}$, and are Kramers degenerate. The separation between levels originating from ${}^2\!E$ and ${}^2T_1$ is exaggerated for clarity. (c): Addition of a weak trigonal perturbative field to the case presented in (b). Irreducible representations $\Gamma_5,\Gamma_6$ are coupled. Note that the ${}^4\!A_2$ ground state is only split (zero field splitting) by a combination of spin-orbit coupling and trigonal distortion, and even then third-order perturbations are required.}
\label{multiplet_levels}
\end{figure}
\begin{figure}[h!]
\includegraphics{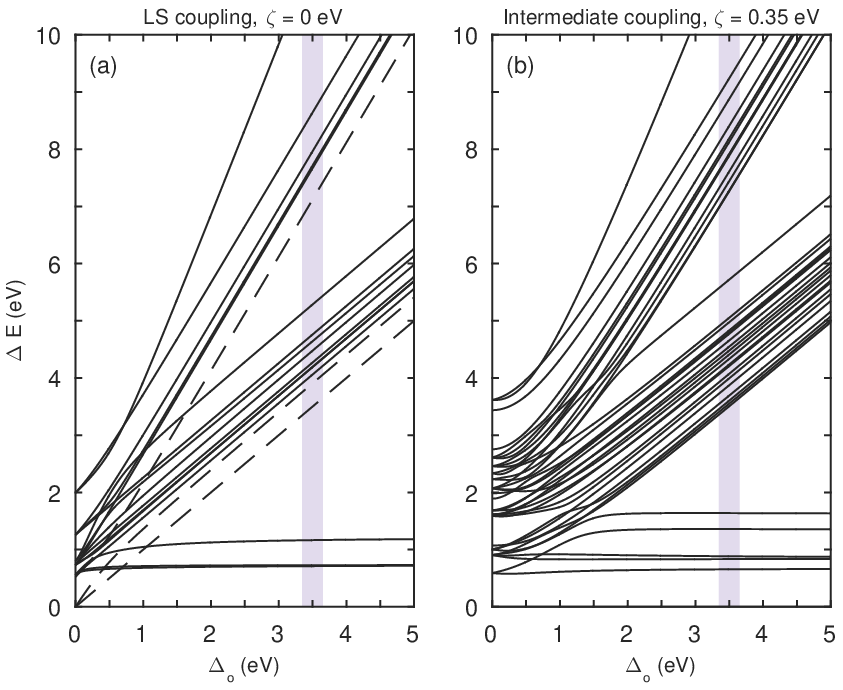}
\caption{Tanabe-Sugano diagrams for octahedral $d^3$ complexes calculated within either (a): the $LS$-coupling limit, or (b): an intermediate coupling model appropriate for \naoso. Trigonal distortion has been neglected for simplicity. Pink line indicates approximate energy levels for \naoso. Dashed lines in (a) indicate spin-allowed ($\Delta S=0$) optical transitions. Both diagrams were calculated using $B=\text{0.035~eV}$, $C=4B$, $J_H = 3B+C = \text{0.245~eV}$.}
\label{TS_d3}
\end{figure}
Yet this is still an oversimplification of the physical picture in \naoso.
For the realistic case of non-zero spin-orbit coupling, and weak trigonal distortion, then peak $\bm{\alpha}$($\bm{\beta}$) should comprise 8(24) overlapping excitations, each with a finite width (Figure \ref{multiplet_levels}).
Clearly it is not possible to resolve -- and fit -- the individual excitations with RIXS, which explains our choice of a simple model to parametrize the data.
 
This in turn explains the observed weak temperature dependence of peak $\bm{\beta}$ (Fig.~\ref{comp_full}d). In a simple crystal field theory, $10\,Dq\propto a^{-5}$, where $a$ is the metal-ligand bond distance. The Os-O bond distance remains practically constant as a function of temperature,\cite{calder2012} therefore $10\,Dq$ should also remain constant. 
Meanwhile $B$ and $C$ are atomic parameters within a ligand field theory. Their magnitude is dictated by the identity and oxidation state of the cation, and the degree of covalence within the metal-ligand bond. These factors should nominally have minimal dependence on temperature. 
The intensity and width of the excitations on the other hand are dictated by the RIXS cross-section, which depends on a wide variety of factors.\cite{ament2011} 
An increase of intensity for the spin-allowed transitions for instance (dashed lines in Fig.~\ref{TS_d3}a) would give rise to an apparent downward shift in the energy of peak $\bm{\beta}$ in our simple model.
Finally, because spin-orbit coupling $(\zeta\sim \text{0.3~eV})$ is of similar magnitude to the inter-electronic interactions ($B,C$) in this material, one should ideally treat them concurrently within an intermediate coupling approach.
Attempts to do this quantitatively for \naoso\ did not to a reliable solution, primarily due to the intrinsically broad excitations. For reference, Figure \ref{TS_d3}b shows a schematic of the respective energy levels expected in a $d^3$ material for sensible values of $J_H$ and $\zeta$.

To summarize, orbital excitations from the ground state do not appear to vary significantly through the MIT. This corroborates the observation that the MIT is not driven by local structural distortion, which would manifest in significant variations in the crystal field parameters $Dq$, $D\tau$ and $D\sigma$.\cite{calder2015_naoso3}
Because these excitations are intrinsically broad, it proved difficult to extract numerical estimates of the Hund's coupling $J_H$, and spin-orbit coupling $\zeta$, in the same manner as presented within Ref.~\onlinecite{taylor2017}.
Nevertheless, by applying a simple $LS$-coupling model with ideal octahedral symmetry, we can estimate $J_H=\text{0.25(1)~eV}$, in good agreement with other studies on 5$d^3$ materials. We note that it may be possible to observe electronic excitations from the ground state via an alternative technique, such as ultraviolet-visible (UV-vis) absorption spectroscopy. This would provide much better energy resolution, at the expense of only being able to measure at the crystallographic zone center.
\\

\section{Low-energy excitations using a high-resolution setup}
\label{lowE_excitations}
We now move to the low-energy excitations below 0.5~eV; the main focus of this paper.
High resolution ($\Delta E=\text{56~meV}$) RIXS spectra were collected at three different momentum transfers as a function of temperature: $\Gamma$ $(\text{4.95},\,\text{2.95},\,\text{3.95})$, $\Gamma$--$Y$ $(\text{5},\,\text{2.75},\,\text{4})$, and $Y$ $(\text{5},\,\text{2.5},\,\text{4})$. This reflects a progression from the Brillouin zone centre to the zone boundary (Fig.~\ref{geometry}d). The point near $\Gamma$ was chosen in order to avoid the weak magnetic Bragg peak at $(\text{5},\,\text{3},\,\text{4})$. 
Four RIXS spectra were collected for each temperature and momentum transfer (30s/pt). 
These spectra were each normalized to the intensity of the intra-$t_{2g}$ excitations at 1~eV energy loss, cross-correlated to account for any temporal shift in the elastic line position, then averaged. 
Measurements were repeated at $\Gamma$ over a limited temperature and energy range; as the elastic line was observed to vary significantly with temperature, partially obscuring the low energy features. We show later that the inelastic features of the two datasets are consistent.\footnote{Furthermore the elastic line intensity was observed to vary significantly between successive scans at $\Gamma$, which was the first point in reciprocal space measured at each temperature. Such behavior was not observed for the other reciprocal lattice points measured. We propose that this may be due to local surface reconstruction or similar effects. Nevertheless, the inelastic features remain consistent between datasets, which means that the following analysis remains valid.}

\begin{figure*}[t!]
\includegraphics{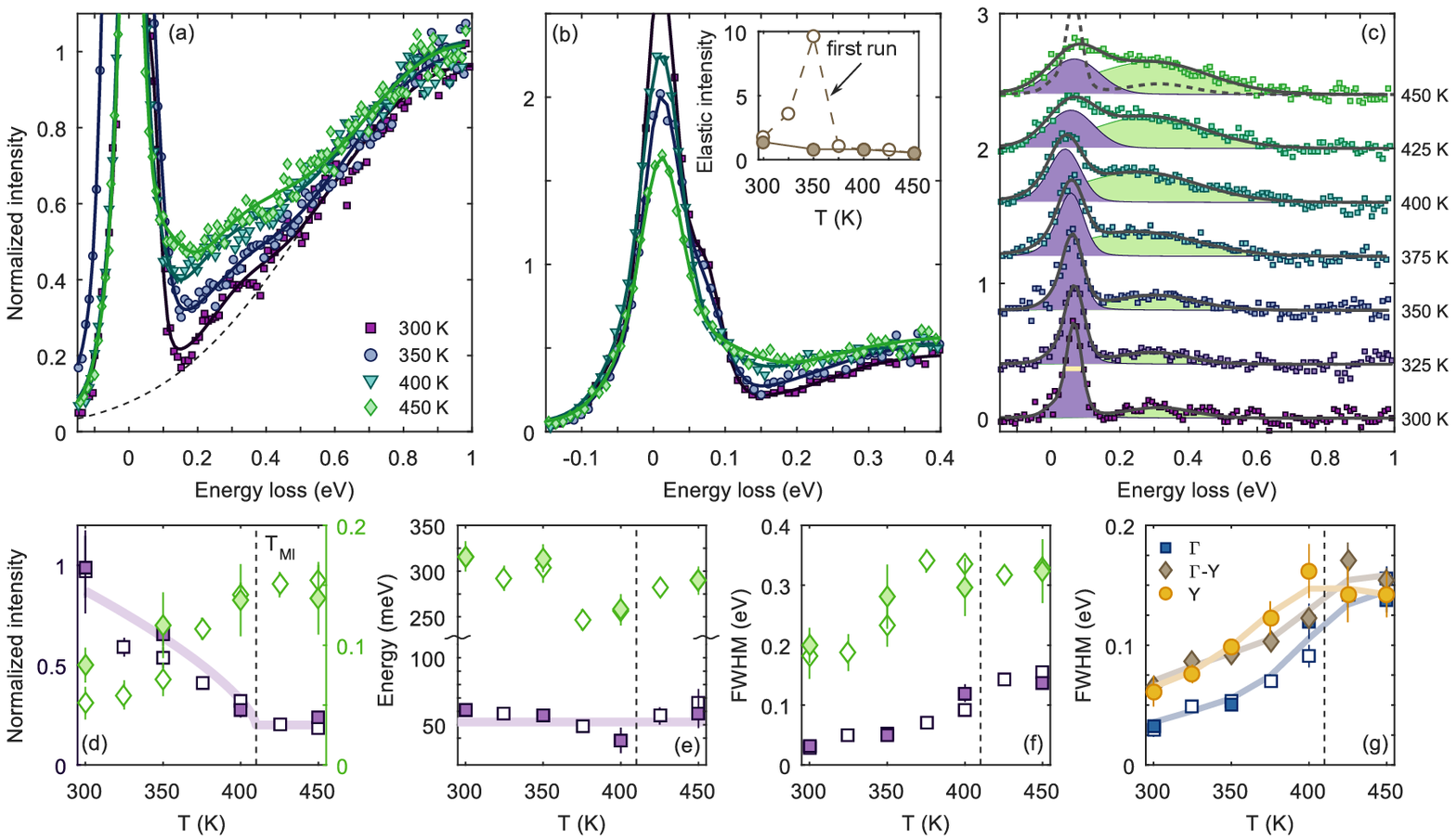}
\caption{Analysis of RIXS spectra collected at $\Gamma$. 
(a,b): Representative RIXS spectra collected in two separate runs, each of which is normalized to the d-d excitations at 1~eV energy loss. The data from the second run (b) was collected over a limited range, and exhibits a significantly reduced elastic line intensity compared to the first run (a). 
Inset in (b) shows the temperature dependence of the elastic line for both runs.
(c): Spectra from first run, with elastic line and d-d contributions subtracted. Added are the best fit to the data (black solid line), and relative components of the magnon peak (purple) and high-energy continuum (green). Dashed line superimposed on 450~K plot is best fit to 300~K data for comparison. Yellow bar indicates full width at half maximum (FWHM) of resolution function.
(d--f): Fitted peak intensity (d), energy (e) and intrinsic FWHM (f) of the two components as a function of temperature. Open symbols: first run. Filled symbols: second run. (g): Summary of magnon peak FWHM as a function of momentum transfer and temperature. Solid lines in (d)--(g) are guides to the eye.}
\label{gamma_fits}
\vspace{12mm}

\includegraphics{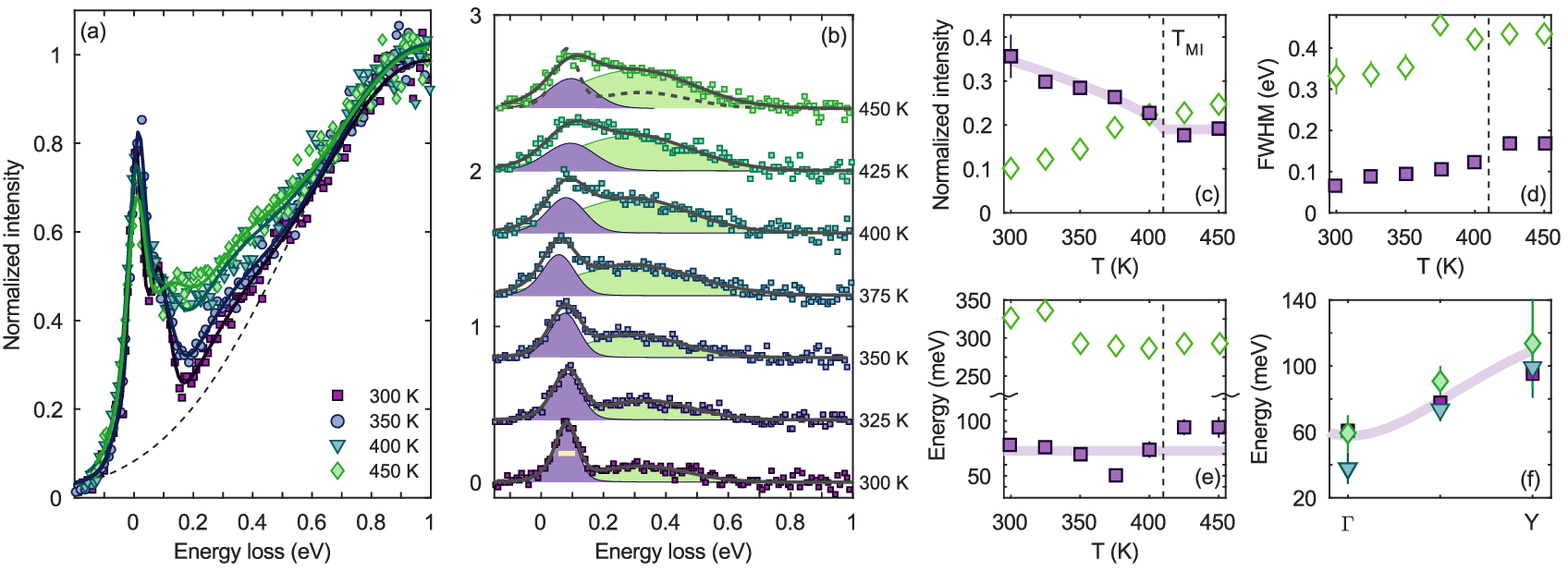}
\caption{Analysis of RIXS spectra collected at $\Gamma$--$Y$. (a): Representative low-energy RIXS spectra, each of which is normalized to the d-d excitations at 1~eV energy loss (dashed line). (b): Spectra with elastic line and d-d contributions subtracted off. Added are the best fit to the data (black solid line), and relative components of the magnon peak (purple) and high-energy continuum (green). Dashed line superimposed on 450~K plot is best fit to 300~K data for comparison. Yellow bar indicates FWHM of resolution function. (c--e): Fitted peak intensity (c), intrinsic FWHM (d), and energy (e) of the two components as a function of temperature. Solid lines are guides to the eye. (f): Energy of magnon peak as function of momentum transfer and temperature. The symbols are the same as part (a). Solid line is best fit to dispersion at 300~K as determined within Ref.~\onlinecite{calder2017_naoso3}.}
\label{gamma_y_fits}
\end{figure*}
\begin{figure*}[t!]
\includegraphics{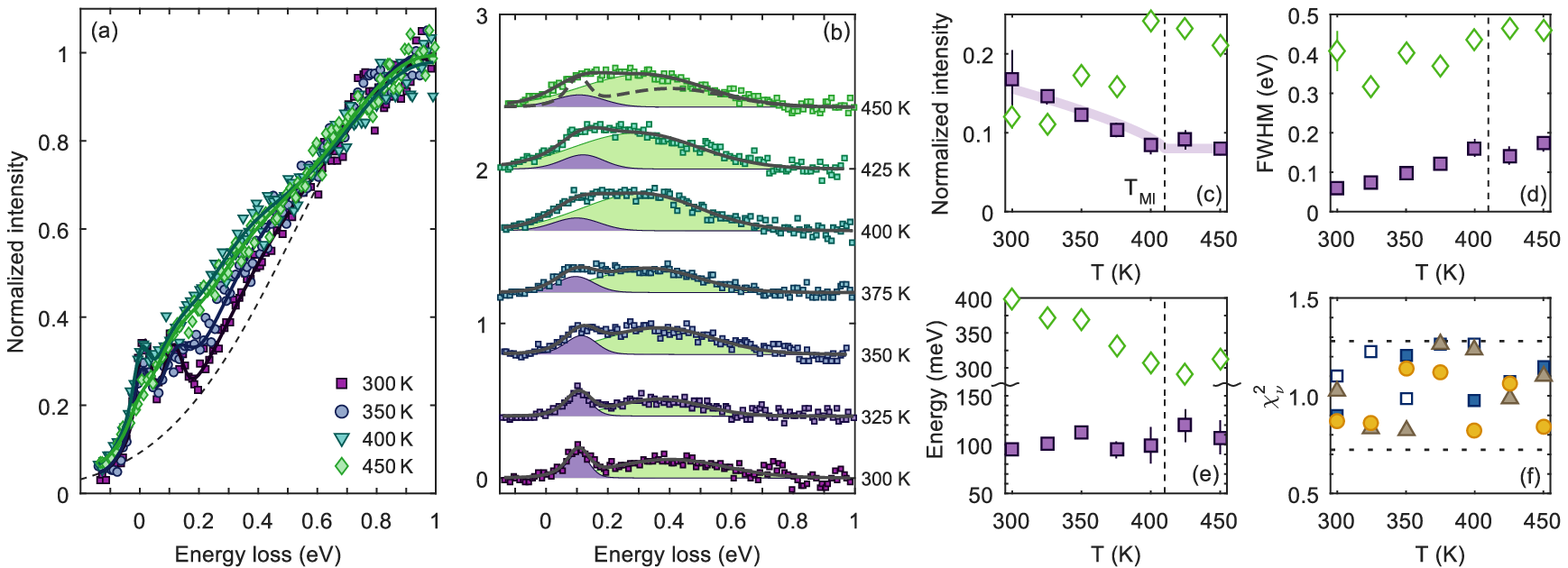}
\caption{Analysis of RIXS spectra collected at $Y$. (a): Representative low-energy RIXS spectra, each of which is normalised to the d-d excitations at 1~eV energy loss (dashed line). (b): Spectra with elastic line and d-d contributions subtracted off. Added are the best fit to the data (black solid line), and relative components of the magnon peak (purple) and high-energy continuum (green). Dashed line superimposed on 450~K plot is best fit to 300~K data for comparison. (c--e): Fitted peak intensity (c), intrinsic FWHM (d), and energy (e) of the two components as a function of temperature. Solid lines are guides to the eye. (f): Values of reduced chi-squared $\chi^2/\nu$ as a function of temperature and momentum transfer. Squares: $\Gamma$, triangles: $\Gamma$-$Y$, circles: $Y$. Dashed lines mark the boundaries of the 95\% confidence limit.}
\label{y_fits}
\end{figure*}


Representative RIXS spectra are plotted in Figs.~\ref{gamma_fits}--\ref{y_fits}a and Fig.~\ref{gamma_fits}b, for temperatures below and above the MIT.
They are also shown with the elastic line and $d$-$d$ contributions subtracted (Figs.~\ref{gamma_fits}c, \ref{gamma_y_fits}--\ref{y_fits}b), in order to better isolate changes to the spectra below 1~eV.
The spectra at 300~K are in agreement with that given in Ref.~\onlinecite{calder2017_naoso3}, with a sharp dispersive peak evident at 60--100~meV energy loss attributable to a single magnon excitation.
With increasing temperature this peak progressively weakens and broadens. Strikingly, concurrent with the diminishing of the single magnon peak, there is a continuous increase in intensity between 0.1 and 0.6~eV, whilst there is no significant change in the intra-$t_{2g}$ excitations.

In order to quantify these observations further, the data were fitted with Gaussians to represent the elastic line, magnon peak, broad component centred around 300~meV, and the intra-$t_{2g}$ excitations. The fits in the low energy portion of the RIXS spectra were corrected to take the Bose factor into account. Prior to fitting, the model lineshape was convoluted with the experimental resolution function (Fig.~\ref{res_ftn}).
This minimal model for the lineshape was used in order to reduce the number of free parameters in the fit, whilst allowing the relevant features of the data to be captured. The results of these fits are given in the remaining panels of Figs.~\ref{gamma_fits}--\ref{y_fits}.
Fig.~\ref{y_fits}f shows that our simple model gives a good description of the experimental data, with $0.8\leq\chi^2/\nu\leq 1.3$ for all datasets.

Clearly there is a significant variation of the RIXS spectra through the MIT, as \naoso\ progresses from the localized to the itinerant limit. The nature of the low-energy excitations at 300~K -- deep in the antiferromagnetic insulating phase -- has already been described in Ref.~\onlinecite{calder2017_naoso3}. Later in this manuscript (Section \ref{intermediate_T}), we perform a more detailed discussion of the temperature dependence of the RIXS spectra. In the meantime, we examine the high temperature behavior; which permits an overview of the interactions towards the itinerant limit, and motivates the analysis of the temperature dependence.

\section{High-temperature behavior (450~K)}\label{HT}
Bulk measurements are typically the first techniques used to understand the electronic and magnetic behavior of a particular system as a function of temperature. \naoso\ is no exception.
Shi \emph{et al.}~found that the high temperature resistivity $\rho$ appears consistent with the theoretical prediction for a Fermi liquid: $\rho = \rho_0 + AT^2$, where $\rho_0$ is the contribution from impurity scattering.\cite{shi2009}
Meanwhile the magnetic susceptibility above $T_{\text{N}}$ exhibits Curie-Weiss behavior, with the Weiss temperature $\mathrm{\Theta_W=-1949\,K}$ indicative of strong antiferromagnetic interactions. Both parameters are plotted in Fig.~\ref{summary} as a function of temperature. These observations suggest that the metallic paramagnetic phase in \naoso\ is rather conventional in nature. 

Various works argue that highly damped spin wave-like excitations, diffusive modes, or both, contribute to the broad magnetic excitations observed in many antiferromagnets above $T_{\text{N}}$.\cite{marshall1968, halperin1969, halperin1976}
The simplest picture is that coherent spin waves, present at low temperature, become progressively more damped as \naoso\ enters the paramagnetic phase. 
Damping occurs primarily due to magnon-magnon scattering, however Landau damping may play a role for itinerant systems. Above $T_{\text{N}}$, these excitations are overdamped. This picture is sometimes known as a ballistic model, referencing principles from kinetic theory.

\begin{figure}[t!]
\centering
\includegraphics{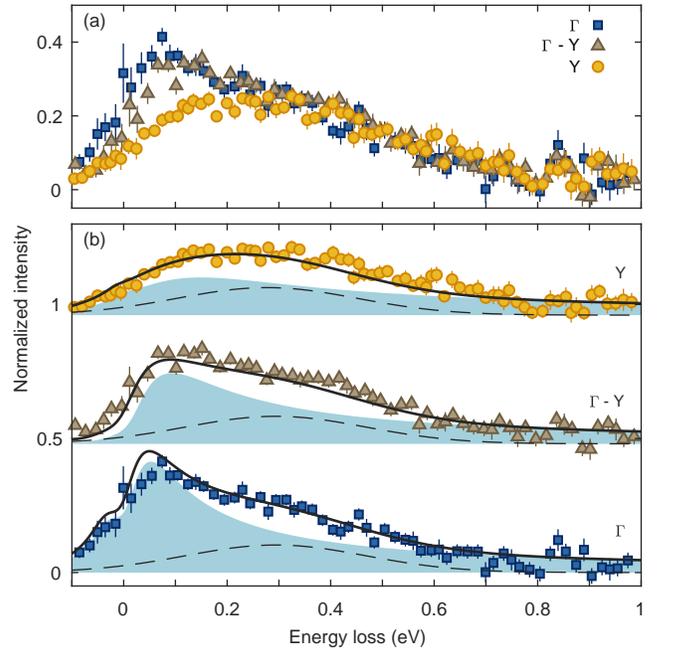}
\caption{Paramagnetic spin fluctuations (SF) in \naoso\ at 450~K. 
(a): Experimental data collected at various crystal momenta. For clarity the elastic line and high-energy intra-$t_{2g}$ excitations have been subtracted. (b): Comparison of WAFL model with experimental data. Filled area indicates paramagnetic SF calculated using Equation \ref{dynamicsusceptibility}, with  $\gamma=0.02~\mathrm{meV}^{-1}$ and $\xi/a_0\sim 1$. Dashed line is a momentum-independent high-energy contribution.}
\label{PM_fluctuations}
\end{figure}

We attempted to fit the data at 450~K (Fig.~\ref{PM_fluctuations}a) using such a model, with the imaginary part of the dynamic susceptibility given by:
\begin{equation}
\chi''(\bm{Q},E) \propto \frac{E}{[\Delta^2 + c^2(\bm{Q}-\bm{Q}_{\text{AFM}})^2-E^2]^2 + \alpha^2E^2},
\end{equation}
where $\Delta$ is the spin gap, $c$ is the spin wave velocity, and $\alpha$ is a damping rate.\cite{tucker2014}. 
As a first approximation, we assumed that the spin wave velocity and spin gap remain constant with temperature. Note however that both of these parameters are likely to decrease in magnitude with increasing temperature, as a consequence of renormalization by quantum fluctuations, magnon-magnon interactions, and other effects.\cite{fishman1998}
We were unable to obtain a satisfactory description of the data with this model, especially at high energies.
This is unsurprising for the following reason. In our simple picture, $\alpha$ should become larger as the temperature increases. The maximum of the spin wave peak shifts to lower energies as a result, concurrent with the build-up of spectral weight on the high energy side.
Yet the experimental data, if anything, shows the opposite trend. We therefore considered alternative models for the excitations in \naoso.

Specifically, we used self-consistent renormalization (SCR) theory appropriate for a weakly antiferromagnetic Fermi liquid (WAFL).\cite{moriya1985} 
The key assumption in the ballistic model is that the paramagnetic state is fully disordered. Yet even though long-ranged order disappears at $T_{\text{N}}$, there may be persistent short-ranged antiferromagnetic correlations, which are characterized by some correlation length $\xi$. These localized clusters of antiferromagnetic order are able to diffuse through the crystal lattice, giving rise to incoherent spin excitations. 

For simplicity we assume that the correlations are spatially isotropic, which is reasonable given the similar magnitude of $J_1$ and $J_2$. Moreover we consider a pseudo-cubic unit cell, with a lattice constant $a_0=3.80~\text{\AA}$ given by the Os-Os distance. This effectively corresponds to the $Pm\bar{3}m$ unit cell of an undistorted perovskite.
Following the approach of Inosov and Tucker, \cite{inosov2010, tucker2014} the imaginary part of the dynamic susceptibility $\chi^{''}(\mathbf{Q}, E)$ is given by:
\begin{equation}\label{dynamicsusceptibility}
\chi^{''}(\mathbf{Q},E) = \frac{\chi_0\Gamma E}{E^2 + \Gamma^2\left[1+\xi^2 (\mathbf{Q}-\mathbf{Q}_{\text{AFM}})^2 \right]^2}.
\end{equation}
In this expression the spin relaxation rate $\Gamma$ is defined through $\Gamma\equiv a_0^2/\gamma\xi^2$, where $\gamma$ denotes the damping coefficient arising from spin decay into particle-hole excitations (related to the electronic band structure), and $\xi$ is the spin-spin correlation length. Furthermore $\chi_0$ is the staggered susceptibility at $\bm{Q}_{\text{AFM}}$.\cite{diallo2010} All of these parameters are in principle dependent upon temperature. The interplay between the correlation length and damping governs the characteristic energy of the damped excitations. 
For completeness the anti-Stokes (energy gain) process has also been included; this is a factor of $\exp{(-\hbar\omega/kT)}$ weaker than the equivalent Stokes (energy loss) process.
Note however that the experimental RIXS cross-section includes a number of additional contributions, which include momentum-dependent absorption and polarization effects. In our analysis we replace the equality in Eqn.~\ref{dynamicsusceptibility} by a proportionality in order to reflect this phenomenologically.

It proved possible to qualitatively describe the main features of the experimental data using this minimal model. Yet quantitative agreement required the use of a scaling parameter that varied by a factor of two across the Brillouin zone. This seemed too large to have a reasonable physical origin.
Inclusion of an additional scattering component improved the quantitative fit to the data, and negated the use of arbitrary scale factors (Fig.~\ref{PM_fluctuations}b). This component was constrained to be momentum-independent, in order to facilitate convergence.
We obtained the best global fit to the data using $\gamma\sim 0.02~\text{meV}^{-1}$ and $\xi/a_0\sim 1$ .\footnote{We note that further improvements in the quantitative agreement with the experimental data could likely be made by considering anisotropic spin fluctuations, or the effect of finite momentum resolution of the spectrometer. However, there are insufficient data to perform this robustly.} The magnitude of $\gamma$ is similar to that found in overdoped Ba(Fe$_{\text{1-x}}$Co$_{\text{x}}$)$_{\text{2}}$As$_{\text{2}}$.\cite{tucker2014} Moreover, the fact that $\xi$ is on the order of the Os-Os separation implies that the spin fluctuations at 450~K are short-ranged, and that the system is close to the hydrodynamic limit.

Thus we conclude that the low-energy excitations at high temperature are consistent with paramagnetic spin fluctuations in a weakly antiferromagnetic Fermi liquid. Such behavior is characteristic of a system close to the itinerant limit, and similarities can be made with magnetic fluctuations in overdoped pnictides. The nature of the additional scattering component centred at 0.3~eV is unclear from the experimental data at 450~K. This is because RIXS is sensitive to both electronic and magnetic interactions, and it is difficult to distinguish them \emph{a priori} without further information. Possible origins are discussed in the following section.

\section{Nature of high-energy continuum}

We considered two likely mechanisms for the formation of this continuum: two-magnon excitations (Section \ref{twomagnon_section}), and electronic interband particle-hole excitations (Section \ref{interband_section}).
These shall be addressed in turn.
\subsection{Two magnon scattering}\label{twomagnon_section}

Consider the simple case of an isotropic 2D Heisenberg antiferromagnet on a square lattice. For a given wavevector, the two magnon continuum extends from the energy of the single magnon, up to twice the zone boundary energy. Notably the spectral weight is concentrated at low energy around the antiferromagnetic Bragg positions at $(\pi,\pi)$, since the single magnon intensity diverges at $(\pi,\pi)$.
Inclusion of anisotropy in the Hamiltonian breaks the rotational degeneracy in the ground state; the single magnon dispersion develops a gap, and the intensity no longer diverges at $(\pi,\pi)$. The two-magnon continuum correspondingly no longer has its lower limit at the one-magnon energy, but also develops a gap. Moreover there is a partial redistribution of spectral weight from the magnetic Brillouin zone centre to the rest of the zone.
In three dimensions the situation is the same, however the scattering phase space is that much greater. This means that spectral weight tends to be concentrated at the zone boundaries, and hence is considerably less than dispersive than the 1D and 2D cases. The two-magnon continuum is a purely quantum effect, and the corresponding dynamic structure factor is typically much smaller than that of the single magnon. \cite{toth2016_LiCrO2}

In a semiclassical picture, single magnons are polarized perpendicular to the magnetic moment direction. Meanwhile the two-magnon excitations are polarized longitudinally to the moment direction.
This manifests as differences in the dynamical correlation function $S^{\alpha\alpha}(\bm{Q},\omega)$, which is defined as
\begin{align}
S^{\alpha\alpha}(\mathbf{Q},\omega)=\frac{1}{2\pi\hbar N} &\sum_{jj'}e^{i\mathbf{Q
}\left(\mathbf{R}_j - \mathbf{R}_{j'}\right)} \nonumber \\ &\times\int^{\infty}_{-\infty} \!\!\! \, \langle S^{\alpha}_{j'} \left(0\right)\!S^{\alpha}_{j}\!\left(t\right)\rangle \,e^{-i\omega t}\,\text{d}t,
\end{align}
where $\alpha$ indexes the Cartesian direction ($x,y,z$) of the spin component, $N$ is the total number of spins, and the sum runs over all sites $j$ and $j'$ in the lattice. Single magnons have nonzero $S^{xx}(\bm{Q},\omega)$ and $S^{yy}(\bm{Q},\omega)$, whereas the two magnon contribution has non-zero $S^{zz}(\bm{Q},\omega)$.
It is well known that the partial differential neutron scattering cross-section is proportional to $S^{\alpha\alpha}(\mathbf{Q},\omega)$.\cite{squires}
Various theoretical works have shown that the RIXS magnetic cross-section resembles $S(\mathbf{Q},\omega)$, at least in the case of the single-band Hubbard model at various filling levels.\cite{haverkort2010, jia2014} This, at least partly, justifies the use of the following approach to estimate the one- and two-magnon scattering cross-sections as a function of energy and momentum transfer. The authors note, however, that the agreement between the RIXS cross-section and $S(\mathbf{Q},\omega)$ may not be as complete in the itinerant limit. Calculation of the RIXS cross-section is numerically involved and remains a subject for future study.

\subsubsection*{Calculation of single- and two-magnon scattering cross-sections}
\newcommand{\bogo}[1]{\ensuremath{{#1}_{\mathbf{q}}}}
\newcommand{\bogoqq}[1]{\ensuremath{{#1}_{\mathbf{q}+\mathbf{Q}}}}
\newcommand{\bose}{\ensuremath{n(\omega_{\mathbf{q}})}}
\newcommand{\boseqq}{\ensuremath{n(\omega_{\mathbf{q}+\mathbf{Q}})}}
\newcommand{\Alad}[2]{\ensuremath{{a}_{#2}^{\text{#1}}}}
\newcommand{\Blad}[2]{\ensuremath{{b}_{#2}^{\text{#1}}}}

The single- and two-magnon scattering cross-sections were evaluated numerically for NaOsO$_{\text{3}}$ using a weighted Monte Carlo method. This approach is similar to that used for the one-dimensional KCuF$_{\text{3}}$, \cite{tennant1995} as well as for the two-dimensional compounds, Cs$_{\text{2}}$CuF$_{\text{4}}$,\cite{coldea2003} and Cu(DCOO)$_{\text{2}}\,\cdot\,$4H$_{\text{2}}$O.\cite{christensen_phdthesis}
We outline the approach used here; further details can be found in Appendix \ref{appendix_mag}. 
\\

Just as in Ref.~\onlinecite{calder2017_naoso3}, the following minimal Hamiltonian was used to model the single magnon mode:
\begin{equation}\label{ham_SM}
\mathcal{H}=J_{\text{1}}\sum_{nn} \mathbf{S}_i\cdot\mathbf{S}_j + J_{\text{2}}\sum_{nnn} \mathbf{S}_i\cdot\mathbf{S}_j + \Gamma\!\sum_{nn,nnn}\!S^z_iS^z_j.
\end{equation}
In Equation \ref{ham_SM}, the first sum is over nearest neighbours (in the $a$-$c$ plane), the second sum is over next-nearest neighbours (in the $b$-direction), and the final term represents an \emph{effective} anisotropy along the $c$-axis. This latter term parametrizes the contributions from single-ion anisotropy or exchange anisotropy (both symmetric and antisymmetric). A full analysis should include all of these individual terms, however it was not possible to disentangle the relative contributions within the experimental energy resolution.
The nearest-neighbour and next-nearest neighbour distances vary only slightly; the difference between them is due to the weak orthorhombic distortion. Thus Equation \ref{ham_SM} is in effect an anisotropic nearest-neighbour Hamiltonian.
From now on we work in a reference frame where the mean spin direction for each site appears along the $z$-axis for all sites. For NaOsO$_{\text{3}}$ this coincides with the laboratory reference frame.

The transverse spin-spin correlations $S^{xx}\left(\mathbf{Q},\omega\right) = S^{yy}\left(\mathbf{Q},\omega\right)$ are given by:
\begin{align}
S^{xx} (\mathbf{Q},\omega) &\propto \frac{S-\Delta S^z}{2}\left[n(\omega_{\mathbf{Q}})+1\right](u_{\mathbf{Q}}-v_{\mathbf{Q}})^2 \nonumber \\
&\times\delta\!\left(\hbar\omega - \hbar\omega_{\mathbf{Q}}\right),
\end{align}
where $n(\omega_{\mathbf{Q}})$ is the Bose factor, $u_{\mathbf{Q}}$ and $v_{\mathbf{Q}}$ are Bogoliubov operators, and $\Delta S^z = S - \langle S^z \rangle$ is the spin reduction due to zero-point fluctuations.\footnote{This is true for the minimal Hamiltonian given by Eqn.~\ref{ham_SM}. However if the Dzyaloshinskii-Moriya (DM) interaction is non-zero, then $S^{xx}\left(\mathbf{Q},\omega\right) \neq S^{yy}\left(\mathbf{Q},\omega\right)$. It was previously noted in Ref.~\onlinecite{calder2017_naoso3} that a finite DM interaction is required by symmetry for NaOsO$_{\text{3}}$.}
In the linear spin wave approximation, the spin reduction can be calculated as: 
\begin{equation} \Delta S^z = (2/N) \sum_{\mathbf{q}} \left[ u^2_{\mathbf{q}} n(\omega_{\mathbf{q}}) \\ + v^2_{\mathbf{q}} (n(\omega_{\mathbf{q}})+1) \right],
\end{equation}
 where $N$ is the total number of spins, and $\mathbf{q}$ extends over the entire Brillouin zone. This expression reduces to $\Delta S = (2/N) \sum_{\mathbf{q}} \left\vert v^2_{\mathbf{q}} \right\vert$ in the limit $T\rightarrow 0$.
For \naoso, we find that $\Delta S^z=\text{0.045}$ at $T=\text{0}$, increasing to  $\Delta S^z=\text{0.081}$ at 450~K. This indicates that quantum fluctuations do not significantly renormalize the spin wave interactions. Such a result is unsurprising given the significant spin wave anisotropy, and three-dimensional nature of this nominally $S=3/2$ system.

Meanwhile the two-magnon cross-section has contributions from three discrete processes: two-magnon creation, two-magnon annihilation, and mixed magnon creation-annihilation. At low temperatures only the two magnon creation term is important; in our case all three are required to give a full description of the data.
The procedure used to calculate the two-magnon cross-section is as follows:
\begin{itemize}
\item Choose two random $\mathbf{Q}$ vectors within the three-dimensional Brillouin zone, henceforth defined as $\mathbf{Q}_1$ and $\mathbf{Q}_2$. These correspond to two separate magnon events.
\item Evaluate the weight of this process $W_{\mathbf{Q}_1, \mathbf{Q}_2}$ using Equation \ref{twomag_weight}.
\item Compute the ratio $w = W_{\mathbf{Q}_1, \mathbf{Q}_2}/W^{\text{max}}_{\mathbf{Q}_1, \mathbf{Q}_2}$, where $W^{\text{max}}_{\mathbf{Q}_1, \mathbf{Q}_2}$ is the weight of the most likely event.
\item Generate a uniformly distributed random number $r$ between 0 and 1, and compare this to $w$. If $w<r$, then the event $(\mathbf{Q}_1, \mathbf{Q}_2)$ is accepted. Otherwise it is rejected.
\item In the case that the event is accepted, then calculate the total energy $\hbar\omega_{\text{tot}}=\hbar\omega_{\mathbf{Q}_1} + \hbar\omega_{\mathbf{Q}_2}$ and total wavevector $\mathbf{Q}_{\text{tot}}=\mathbf{Q}_1+\mathbf{Q}_2$. If $\mathbf{Q}_{\text{tot}}$ lies outside the first Brillouin zone, then it is folded back via subtraction of a reciprocal lattice vector $\mathbf{G}$.
\end{itemize}
This process is repeated until a large number of events are accepted.
For NaOsO$_{\text{3}}$, $1\times 10^{9}$ iterations were performed in order to generate $S^{xx(yy)}(\mathbf{Q},\omega)$, and the inelastic contribution to $S^{zz}(\mathbf{Q},\omega)$ as a function of momentum transfer and energy. These functions were then normalised by $(S-\Delta S^z)(2\Delta S^z+1)$ and $\Delta S^z(1-\Delta S^z) + \langle uv\rangle^2$ respectively, in order to obtain a direct comparison with experiment. 

The experimental geometry governs the relative contributions of $S^{xx}(\mathbf{Q},\omega)$, $S^{yy}(\mathbf{Q},\omega)$, and $S^{zz}(\mathbf{Q},\omega)$ to the scattering cross-section. It is reiterated that this may not be equal to the RIXS cross-section, as orbital effects for example are neglected.
Haverkort \cite{haverkort2010} proposed a simple selection rule for the measurement of single spin-flip excitations (magnons) with RIXS: Magnons can be observed with cross-polarised light for spins in the plane of the polarizations. 
We have almost entirely $\pi$-incident polarization (undulator source), which means that the incoming X-rays are polarized parallel to the scattering plane. The cross-polarized channel lies perpendicular to the scattering plane, given the notation $\sigma$ in line with current convention. Consequently we observe those spin components which lie within the plane defined by the $\pi$ and $\sigma'$ polarization vectors, with the prime denoting the polarization of the scattered beam (Fig.~\ref{geometry}a).
This contrasts with the selection rule for magnetic neutron scattering, in which only magnetic fluctuations perpendicular to the wave vector transfer $\bm{Q}$ are observed.
For the present scattering geometry, the intensity is hence approximately given by the relation:
\begin{align}
I \propto S^{yy}(\bm{Q},\omega)\cos{\theta} + \sqrt{2}\sin{\theta}\left[S^{xx}(\bm{Q},\omega)+S^{zz}(\bm{Q},\omega)\right],
\end{align}
where the angle $\theta$ is defined by the rotation of the sample within the scattering plane. A small momentum-dependent tilt out of the scattering plane by $<\!\text{4}^{\circ}$ has been neglected; this contributes only weakly to the intensity and does not significantly affect the results presented here.

\begin{figure}[t!]
\includegraphics{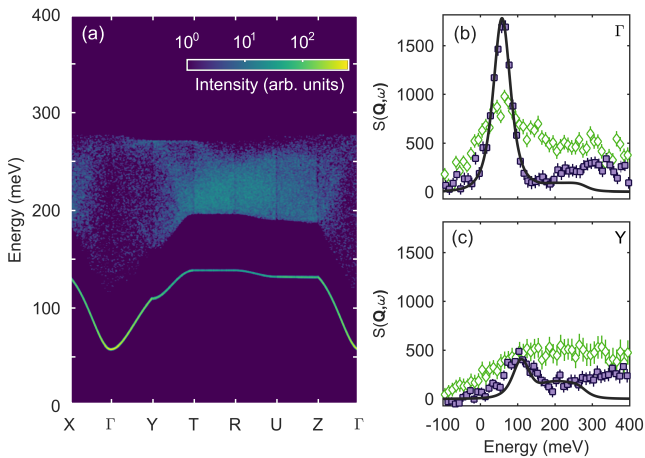}
\caption{(a): Transverse $\left[S^{xx}(\mathbf{Q},\omega)\right]$ and longitudinal $\left[S^{zz}(\mathbf{Q},\omega)\right]$ spin-spin correlation functions, calculated as a function of energy and momentum transfer. The simulation ran over $1\times10^{9}$ Monte Carlo steps, with each displayed single and two-magnon event broadened by a two-dimensional Gaussian with FWHMs 0.5~meV and 0.01 reciprocal lattice units (r.~l.~u.) for clarity. No spin wave damping has been assumed. All Bose factors have been calculated assuming $T=\text{300~K}$. Lines at high symmetry points are artefacts of the plotting routine.
(b, c): Comparison of the $S(\mathbf{Q},\omega)$ calculations (solid line) with uniformly scaled experimental RIXS data at $\Gamma$ (b) and $Y$ (c). Filled squares refer to data obtained at 300~K, open diamonds to data obtained at 450~K.  The RIXS data has had the elastic and $d-d$ contributions subtracted, in order to highlight the remaining features. The calculated $S(\mathbf{Q},\omega)$ has been convoluted with a Pearson VII function of FWHM 56~meV and profile parameter $\mu=2$, in order to represent the effect of the instrumental resolution. Note that this does not include the effect of the finite momentum resolution of the spectrometer.}
\label{onetwo_magnon}
\end{figure}
\subsubsection*{Results}
The results of this simulation are plotted in Fig.~\ref{onetwo_magnon}a, assuming no spin wave damping. A clear dispersive mode can be observed between 60 and 130~meV, along with a broad continuum of states which extends up to 260~meV. These correspond to the single and two-magnon excitations respectively. Anisotropy not only induces a gap in the single magnon dispersion, but also acts to separate the single and two-magnon excitations.
From Fig.~\ref{onetwo_magnon}a, it appears that the two magnon continuum is most concentrated at the zone boundaries, as expected.
Clearly the simulations qualitatively reproduce the expected behavior for both the single magnon and two magnon continuum. Quantitative comparisons are provided in Figs.~\ref{onetwo_magnon}b,c at $\Gamma$ and $Y$ respectively. The relative intensities of the single magnon peak at 300~K, and different momentum transfers, are well described by the Monte Carlo simulation. 

On the other hand, the high energy spectral weight cannot be fully modelled in terms of this simple model for the two-magnon continuum. Furthermore, the quantitative agreement at 450~K is not as good, even when taking the difference in the Bose factor into account.
There are, however, two factors which have not been considered thus far.
The first is that the calculations assume a purely localized model (Heisenberg), and do not include the effect of magnon damping. 
Landau damping is likely to occur for magnons with energies larger than the electronic charge gap ($E_g \sim \text{80~meV}$ at 300~K).
This hypothesis is partly justified by the observation that the Monte Carlo simulations seem to slightly underestimate the experimental peak width of the zone boundary ($Y$) single magnon peak at 100~meV. Such an effect would also likely apply to the two-magnon excitations.

The second factor is that the experimental RIXS spectra are the average over all outgoing polarization channels. This means that they not only include contributions from the cross-polarised $\pi$--$\sigma'$ channel, but also from the $\pi$--$\pi'$ channel. The $\pi$--$\pi'$ channel contains all non-magnetic scattering components, including electronic excitations. If the charge gap is small, then one may observe charge scattering from a broad continuum of (weakly momentum-dependent) interband particle-hole excitations. Such an effect has been previously observed via RIXS in Na$_{\text{2}}$IrO$_{\text{3}}$, and may partly be the origin for the high-energy scattering beyond 0.3~eV.\citep{gretarsson2013_prl, gretarsson2013_na213, bhkim_2014} Note that any excitonic behavior (if present) is likely to be hidden by the single-magnon peak at a similar energy.

\subsection{Inter-band transitions}\label{interband_section}

\begin{figure}
\includegraphics{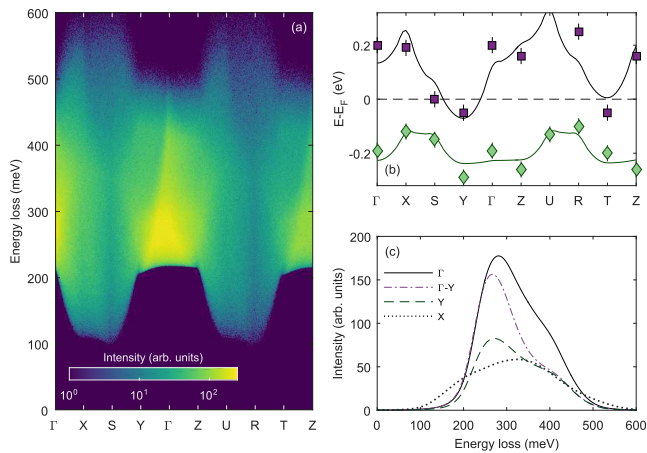}
\caption{Calculation of dipole-allowed interband transitions for \naoso. (a): Results of Monte Carlo simulation ($5\times10^{8}$ events) shown for a portion of the Brillouin zone. (b): Theoretical band structure of \naoso\ at $T_{\text{N}}$, as calculated within Ref.~\onlinecite{bongjaekim2016}. Symbols indicate values at high symmetry points extracted from manuscript, with error bars reflecting the uncertainty in this procedure. Solid lines are results of fits to this band structure within the parabolic approximation. The main features of the band structure are reproduced. Fitting parameters $\left[m_x^*, m_y^*, m_z^*, E_0\right]$ for conduction band: [0.8(2), 2.4(3), 0.4(2), $-$0.07(3)~eV]; for valence band: [0.6(2), 0.1(3), 0.0(2), $-$0.24(3)~eV]. All effective masses are given in units of $m_e$. (c): Cuts of data presented in (a), taken at various reciprocal lattice points as indicated. The effect of the instrumental energy resolution has been included.}
\label{NaOsO3_interband}
\end{figure}

In order to test the validity of this prediction, we performed calculations of potential dipole-allowed inter-band transitions for \naoso. The starting point was the band structure calculated by Bongjae Kim and colleagues at $T_{\text{N}}$ (Fig.~3c in Ref.~\onlinecite{bongjaekim2016}).
To simplify matters we only consider the two bands closest to the Fermi energy, which were fitted within the parabolic band approximation:
\begin{align}
E(\mathbf{k})=E_0+\frac{\hbar}{2}\left[\frac{(k_x-k_{x,0})^2}{{m_x^*}} + \frac{(k_y-k_{y,0})^2}{{m_y^*}} \right. \nonumber \\ + \left.\frac{(k_z-k_{z,0})^2}{{m_z^*}}\right]
\end{align}
where $m^*_{x,y,z}$ refers to the effective mass in the respective Cartesian direction, and $E_0$ is the energy at the band minimum located at $\left(k_{x,0},k_{y,0},k_{z,0}\right)$. This was assumed to be at $Y$. The band energies at high symmetry points (as calculated by Kim) were used to constrain the fits. The results of this fit are displayed in Fig.~\ref{NaOsO3_interband}b, with all the main features of the two electronic bands reasonably reproduced.

Fig.~\ref{NaOsO3_interband}a shows the results of a Monte Carlo simulation performed over $5\times10^{8}$ events. The general method was similar to that used to calculate the one- and two-magnon cross-section. A broad, weakly momentum dependent, continuum can be observed around 0.3~eV energy loss, which is of a similar scale to the high energy peak observed in the RIXS data (Fig.~\ref{NaOsO3_interband}c). This fits with the hypothesis that there may indeed be an intrinsic electronic component present within the RIXS data.

However, this conclusion suffers from a number of limitations. 
Fundamentally, RIXS at the L$_{\text{3}}$ edge is not a direct probe of the band structure, even in weakly correlated systems. We note that K-edge (indirect) RIXS can measure the band structure through the joint density of states (JDOS), with a more detailed discussion given in Refs.~\onlinecite{hu2008} and \onlinecite{kim2010_cu2o}. Unfortunately the Os K-edge lies at 73.9~keV, which is well above the present capabilities of any currently available RIXS instrumentation. 
Secondly we only considered transitions between the two bands closest to the Fermi level; there are likely to be additional contributions to the observed scattering from neighbouring bands.\footnote{We note that a recent paper by Mohapatra \emph{et al.}~parametrises the electronic band structure within a three-orbital tight-binding model\cite{mohapatra2017}} Moreover the calculated band structure has not been verified experimentally -- for example via photoelectron emission spectroscopy (PES)-- and contains a degree of fine structure which has been averaged out within the present analysis.

The calculated amplitude of the interband transitions appears to decrease by a factor of two at the zone boundary ($Y$), compared to the zone centre ($\Gamma$). This is not observed in the RIXS data (compare with Figs.~\ref{gamma_fits}c and \ref{y_fits}c), however note that the finite momentum resolution of the spectrometer has not been taken into account when performing the interband calculations. There are also likely $\bm{Q}$-dependent geometrical, absorption, and orbital effects, which enter the RIXS cross-section that are not considered here.
Finally the energy scale of the continuum is considerably larger than $kT_{\text{N}}$. We hence expect the scattering intensity to vary little with temperature, provided that the temperature change is not too large. This is at apparent odds with the experimentally observed behavior. 

To summarise, the high energy continuum likely contains two components: two-magnon scattering, and interband particle-hole excitations. We posit that it is dominated by the latter, even at 300~K, since two-magnon scattering alone cannot describe the high energy ($>\text{0.4~eV}$) portion of the continuum.
These interband transitions appear to be weakly dependent upon temperature and momentum transfer, which makes sense given that \naoso\ undergoes a continuous MIT.
They cannot, however, explain the large renormalization of RIXS intensity with temperature. 

\section{Magnetic excitations at intermediate temperatures}\label{intermediate_T}
Thus far the focus has been upon measurements performed close to the localized and itinerant limits (at 300~K and 450~K respectively). The question is, what happens to the magnetic excitations at intermediate temperatures? From Figs.~\ref{gamma_fits}--\ref{y_fits}, it appears that the low-energy excitations continuously evolve through the MIT.
The experimental RIXS spectra were analysed using the simple fitting model presented in Section \ref{lowE_excitations}.
We note at this point that this model (and the one given in Ref.~\onlinecite{calder2017_naoso3}) assumes that the magnons are underdamped, whereas it has already been shown that the behaviour at 450~K appears consistent with overdamped spin excitations (Section \ref{HT}). However the ballistic and diffusive regimes are not mutually exclusive; there is likely to be a crossover between the two (with potential phase coexistence) in the vicinity of the MIT. For simplicity we do not consider this in our analysis. 
Our discussion shall focus upon the temperature dependence of the magnon peak. 

As expected, the single magnon peak appears to weaken progressively with temperature for all momentum transfers (Figs.~\ref{gamma_fits}d, \ref{gamma_y_fits}--\ref{y_fits}c). This abates at $T_{\text{N}}$, with some residual intensity remaining all the way to 450~K.
The temperature dependence is consistent with a power law, again as expected for a magnetic excitation. Due to limited data, it is not possible to obtain a reliable value for the critical exponent. We note the following.
In Ref.~\onlinecite{calder2012}, the authors proposed that $\beta \approx 0.3(1)$, based on fits of the magnetic scattering (from neutron powder diffraction) over an extended temperature range. This value is consistent with three-dimensional magnetic ordering. Yet close to $T_{\text{N}}$ -- where the power law approximation is more reliable -- a distinct linear regime can be observed within the sizeable experimental uncertainty.
In an ideal Slater picture, one treats the magnetic interactions within a mean field. Therefore the critical exponents would be expected to follow a Ginzburg-Landau model, in which $\beta = 1/2$. In resonant (in)elastic x-ray scattering, the magnetic scattering cross-section below $T_{\text{N}}$ is proportional to $M^{-2\beta}$, where $M$ is the magnetization. Hence the magnetic scattering intensity should decrease linearly with temperature.
Our data is broadly consistent with either scenario.
More precise measurements are required to ascertain the true dimensionality of magnetic interactions in the vicinity of the MIT.

A weak variation in the energy of the magnon peak with temperature can be observed. At $\Gamma$, the magnon appears to soften by 26(9)~meV between 300~K and 400~K (Fig.~\ref{gamma_fits}e). A similar value is obtained at $\Gamma$-$Y$ between 300~K and 375~K [28(4)~meV, Fig.~\ref{gamma_y_fits}e]. However, the magnon at the zone boundary remains unperturbed in the same temperature range (Fig.~\ref{y_fits}e). These observations are summarized in Fig.~{\ref{gamma_y_fits}}f, which also shows that the spin wave dispersion at 450~K is remarkably similar to that collected at 300~K (within experimental resolution). 

We make the following remarks.
Firstly the magnitude of the anisotropy presented in Ref.~\onlinecite{calder2017_naoso3} is likely to be an overestimate. This is a direct result of the finite momentum resolution of the RIXS spectrometer, which most prominently affects the magnetic excitations at the Brillouin zone centre, where the dispersion is steepest.\cite{pincini2017} We estimate the in-plane momentum resolution $\Delta Q \approx 0.17\mathrm{\AA}^{-1}$ (at FWHM). Combined with the finite energy resolution, this means that any temperature dependence of the spin gap is likely to be washed out and very difficult to observe. We discuss this point further in Appendix \ref{resolution_ftn}.
Secondly a reduction of the spin gap with increasing temperature, or renormalization of the exchange parameters due to magnon-magnon interactions, would lead to a uniform softening throughout the Brillouin zone.\footnote{K{\"{o}}bler \emph{et al.} predict that the spin wave stiffness, optical magnon, and order parameter should have identical temperature dependences in magnets with quenched orbital moments.\cite{kobler2005} For instance, they determine that these parameters should scale with $T^{3/2}$ for 3D anisotropic magnets.} This is not observed within experimental uncertainty.
Furthermore, the bulk of the observed softening is restricted to one temperature step. Hence we cannot state definitively whether our observations correspond to a real physical effect.

Finally the width of the magnon peak increases as a function of temperature for all momentum transfers. This is unsurprising given that there a number of mechanisms for magnon damping to occur.
These can broadly speaking be split into those which apply in the localized and itinerant limits.
The dominant mechanism for magnon damping in localized (Heisenberg) magnets is four-magnon (two-magnon in, two-magnon out) scattering, and approximately scales as $qT^4$.\cite{cottam1970, stinchcombe1974, bayrakci2013}  
Magnon-phonon coupling has also theoretically been shown to be a source of damping in the localized limit.\cite{cottam1974}

In itinerant magnets, there is a fundamental coupling between the spin, electronic, and lattice degrees of freedom. Spin wave excitations are damped within the Stoner continuum due to scattering from intraband particle-hole excitations (Landau damping).
Within a weak coupling theory,\cite{feddersmartin_1966} collective antiferromagnetic spin wave excitations are expected to merge into a Stoner continuum above a critical energy 
\begin{equation}\label{Stoner_eqn}
\Delta_{\text{s}}\!~\approx~\!\pi k_{\text{B}} T_{\text{N}}\left[8\alpha^2 t/7\zeta(3)\right]^{1/2}
\end{equation} 
where $T_{\text{N}}=\text{410~K}$ is the N\'{e}el temperature, $t=1-T/T_{\text{N}}$, $\alpha=\nu_e\nu_h/\sqrt{4\nu_e\nu_h}\approx 1$ is a dimensionless parameter relating the electron and hole band velocities, and $\zeta(3)$ is the Riemann zeta function of the third kind. 
Note the mean-field temperature dependence. As $T\rightarrow T_{\text{N}}$, the energy scale for $\Delta_s$ decreases, meaning that the collective excitations become damped over a larger proportion of the Brillouin zone. This argument is expected to remain valid for weak electronic correlations $U$.
Meanwhile itinerant electrons concurrently give rise to overdamped collective modes, which are the aforementioned spin fluctuations.\cite{moriya1985} Scattering from these can also act as a further mechanism for magnon decay.\cite{solontsov1993}

In Ref.~\onlinecite{calder2017_naoso3}, it was proposed that the well-defined, resolution limited spin wave excitations at 300~K were evidence of localized (Heisenberg) behavior.
Yet with the aid of our new fitting model, we observe experimentally that the magnons have an intrinsic FWHM of 30~meV, increasing to 60~meV at larger momentum transfers. Some of this damping may arise from the four-magnon, or magnon-phonon scattering processes mentioned earlier. 
There is, however, the omnipresent continuous MIT at 410~K. Previous bulk measurements have shown that the charge gap $\Delta_g \sim \text{80~meV}$ at 300~K,\cite{shi2009} with the optical gap of a similar magnitude.\cite{lovecchio2013} Returning to Eqn.~\ref{Stoner_eqn}, and taking $\alpha\!=\!1$, then we find that $\Delta_{\text{s}}\,[T=\text{300~K}]\approx \text{60~meV}$.
This would apply that the collective spin excitations should be (weakly) Landau damped at all wavevectors.
At higher temperatures, $\Delta_s$ decreases concomitantly with the charge and optical gaps. This leads to a greater tendency towards Landau damping. As mentioned earlier, emergent SF due to itinerant electrons can give rise to additional magnon decay channels. We tentatively suggest that both mechanisms may be enhanced by the presence of magnon-phonon coupling.
\\

To summarise, we find that the temperature dependence of the single-magnon peak is consistent with that expected for a magnetic excitation. Damping increases as the system progresses towards the metallic state, due to Landau damping and scattering from itinerant spin fluctuations. The role of magnon-phonon coupling on the RIXS spectra is still unclear. We posit that it may result in a downward shift of the magnon dispersion of around 5~meV at the zone boundary; a similar magnitude to that observed for the phonon shift.

\section{Discussion and summary}
In summary, we find that at 300~K, \naoso\ lies close to, but not within, the Heisenberg limit. The magnetic scattering is dominated by single-magnon processes, albeit with some contribution from two-magnon and inter-band particle-hole excitations. Spin-phonon coupling may play a role in perturbing the magnon dispersion at the zone boundary, however it is difficult to be certain given the experimental resolution.
The previously reported magnitude for the effective anisotropy $\Gamma=\text{1.4(1)~meV}$ -- containing contributions from single-ion and exchange terms -- is likely to be an overestimate of the true value due to resolution effects. 
Given that anisotropy (single-ion or exchange) is governed primarily by spin-orbit coupling (SOC), a lower value of $\Gamma$ would imply that SOC plays less of a role in the formation of the magnetic ground state and excitations than previously believed.

With increasing temperature, one observes a continuous progression towards the itinerant limit through the MIT. At 450~K, significant paramagnetic spin fluctuations -- consistent with a weakly antiferromagnetic Fermi liquid model -- are present, along with an approximately momentum-independent additional component. Similarities can be drawn with properties seen in the doped cuprates and iron pnictides as a function of temperature and carrier doping.\cite{letacon2011, dean2013, monney2016, diallo2009, zhao2009, zhou2013, wang2013, leong2014, dai2015}
Future measurements using a different experimental geometry (grazing emission for instance), or polarization analysis of the scattered beam, may help to disentangle the spin and electronic contributions to the RIXS spectra. 

Our observations directly map themselves onto the discussion of the proposed Slater MIT in this material. Recall that insulating behavior in Slater-type materials arises from an ordered magnetic exchange field, which is governed by mean-field type interactions. Previous RIXS measurements on other $5d^3$ materials have shown that the orbital excitations can only be described by an intermediate-coupling model in which SOC plays a key role.\cite{taylor2017} Our results for \naoso\ (Section \ref{orbital_excitations}) are compatible with this picture. Furthermore, the use of the WAFL model to describe the paramagnetic spin fluctuations naturally requires the presence of correlations in the high temperature phase; which implies a departure from mean-field behavior required for a pure Slater MIT. 
Therefore we conclude that \naoso\ lies proximate to, but not within, the Slater limit.


Finally we show that RIXS opens a new window on the progression of quasi-particle spectra through metal-insulator transitions. Unlike the cuprates and pnictides, there is a continuous evolution between the localized and itinerant limits in the same sample of \naoso. This means that any differences in the RIXS spectra between the two limits are intrinsic, and not a consequence of disorder, variation in measurement conditions, or other perturbative effects.
Our results show that functional forms applicable to inelastic neutron scattering from itinerant magnets are also semi-quantitatively valid in RIXS. Whilst further theoretical work is required to confirm this, it confirms the correspondence between $S(\bm{Q},\omega)$ and the RIXS cross-section in the itinerant limit.


\acknowledgments{J.~G.~V. thanks University College London (UCL) and \'{E}cole Polytechnique F\'{e}d\'{e}rale de Lausanne (EPFL) for financial support through a UCL Impact award, and useful discussions with B. J. Blackburn, A. Princep, and E. V\"{a}is\"{a}nen. Work at UCL was supported by the EPSRC (grants EP/N027671/1, EP/N034872/1).  This research used resources at the High Flux Isotope Reactor and Spallation Neutron Source, DOE Office of Science User Facilities operated by the Oak Ridge National Laboratory. K.Y. thanks financial support from JSPS KAKENHI (15K14133 and 16H04501). All data created during this research are openly available from the UCL Discovery data archive at \url{http://dx.doi.org/10.14324/000.ds.1550187}.}

\appendix

\setcounter{figure}{0}
\renewcommand\thefigure{\thesection\arabic{figure}}   

\section{Spectrometer resolution function}\label{resolution_ftn}
\begin{figure}[ht]
\centering
\includegraphics{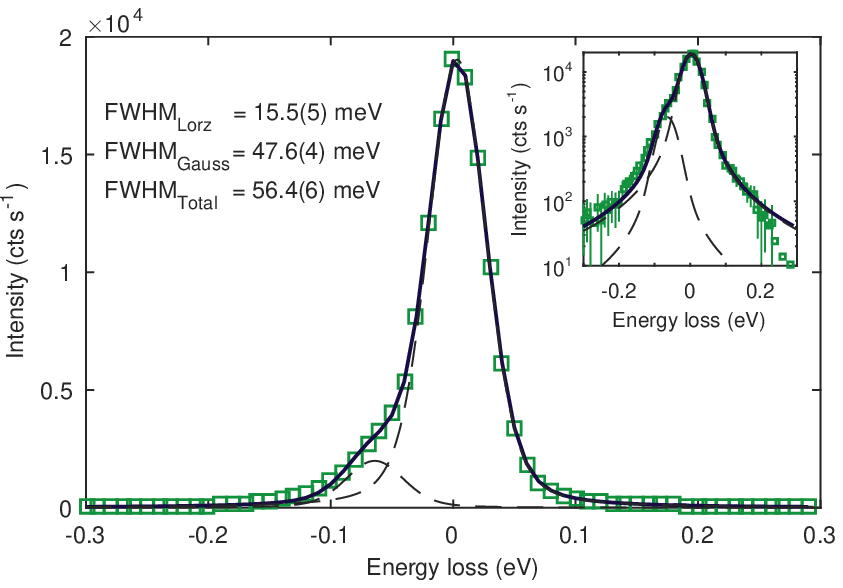}
\caption{Spectrometer resolution function for \naoso\ obtained from diffuse scattering off transparent adhesive tape. Solid line is best fit to sum of two Voigt functions with the same Lorentzian and Gaussian widths. The asymmetry arises due to the fact that $2\theta_A$ is significantly away from $90^{\circ}$ at the Os L$_{\text{3}}$ edge. Inset is the same as the main panel, only on logarithmic axes to highlight the peak tails. This profile was used to fit all of the experimental data.}
\label{res_ftn}
\end{figure}
The effect of the finite energy resolution of a RIXS instrument operating in the hard x-ray regime is well known. The intrinsic width (in energy) of an arbitrary excitation is proportional to the inverse lifetime of the core-hole in the final state. This contrasts with x-ray absorption spectroscopy (XAS), in which the spectra are broadened by the core-hole in the intermediate state.
What we observe experimentally is a convolution of this excitation with the spectrometer resolution function.
The dominant contributions to the resolution function include the monochromator configuration, vertical divergence of the incident beam, choice of analyser crystal and the available collection area for scattered photons. A number of tables exist which allow the calculation of the expected energy resolution for a given experimental configuration.\cite{analyser_atlas}

In Fig.~\ref{res_ftn}, we show the energy component of the instrumental resolution function.
The total energy resolution was determined to be $\Delta E = \text{56~meV}$, based on diffuse scattering from polypropylene-based adhesive tape.
This is marginally worse than both the resolution obtained in Ref.~\onlinecite{calder2017_naoso3} ($\Delta E =\text{45~meV}$), and the theoretical value for our given setup ($\Delta E_{\text{calc}}=\text{42~meV}$). The difference is likely to be related to a slight difference in instrumental configuration.

However there has been little consideration of the finite \emph{momentum} resolution of the spectrometer in the literature. This is in contrast with inelastic neutron scattering, where this is routinely performed.
This may be partly due to historical reasons with the development of RIXS. A large portion of RIXS performed thus far has been at the 3d transition metal $L$-edges, which lie in the soft x-ray region. Spectrometers operating in this window are based on diffraction gratings, and typically have in-plane momentum resolutions $\Delta Q_{||}\sim 10^{-2}\,\mathrm{\AA}^{-1}$. This is sufficiently small -- especially compared to the energy resolution -- that the effect of the finite momentum resolution of the spectrometer is negligible for all intents and purposes.

Meanwhile hard x-ray RIXS instruments utilize crystal optics and much higher energy incident radiation. Whilst it has the advantage that multiple Brillouin zones can be accessed and sampled, it comes with the drawback of momentum resolution around an order of magnitude worse than in the soft x-ray regime. The dominant contribution to $\Delta Q_{||}$ tends to be the radius of the crystal analyser $r_{an}$. Reducing $r_{an}$ improves $\Delta Q_{||}$ at the expense of lower count rates. 
Note that the momentum resolution ellipsoid is -- like a neutron triple-axis instrument -- three-dimensional and anisotropic. Therefore the precise momentum resolution at a given point in reciprocal space depends on the experimental geometry.

\begin{figure}[t]
\includegraphics{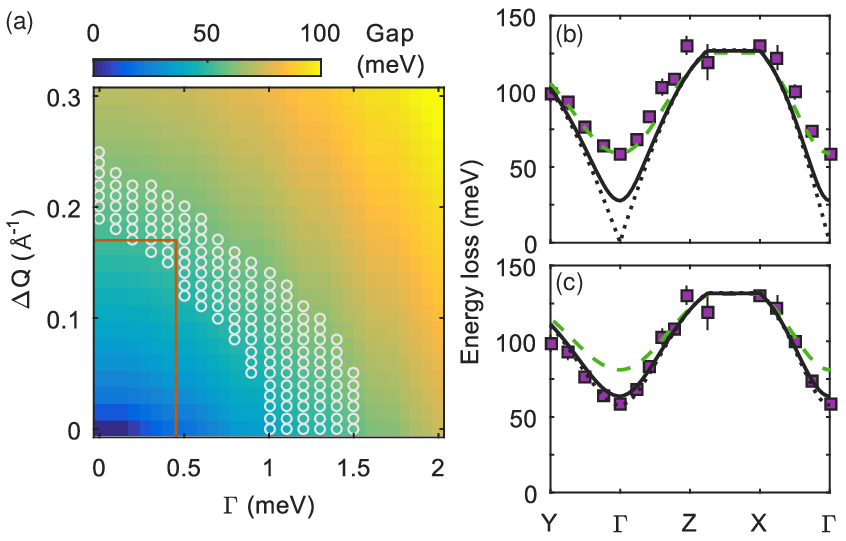}
\caption{Effect of finite momentum resolution $\Delta Q$ on the magnon dispersion. (a): Calculated magnon gap at $\Gamma$ assuming the Hamiltonian given by Eqn.~\ref{ham_SM}, and $J_1=J_2=\text{14~meV}$. Open circles reflect values of $\Delta Q$ and $\Gamma$ which are consistent with the experimentally observed gap of 55(7)~meV. Solid line shows $\Gamma=\text{0.5(3)~meV}$ for $\Delta Q=\text{0.17 \AA}^{-1}$. (b,c): Comparison of experimental data with simulated dispersions broadened by $\Delta Q = \text{0, 0.1, 0.2}\,\text{\AA}^{-1}$ (dotted black, solid black, dashed green lines respectively). For (b), $\Gamma=\text{0}$, while $\Gamma=\text{1.4~meV}$ for (c).}
\label{Qres}
\end{figure}  

Non-zero momentum resolution will lead to the smearing of the magnon dispersion. This will have the greatest effect where the spin-wave velocity is steepest; typically around the crystallographic and magnetic Brillouin centres. In the most extreme case, it will lead to an apparent gap being observed, even when considering a theoretical model without anisotropy. More generally, the magnitude of any anisotropy present in the system will be overestimated, if the effect of momentum resolution is not considered.

Previous fitting of the magnetic excitations in \naoso\ assumed perfect momentum resolution, and led to an estimate for $\Gamma=\text{1.4(1)~meV}$. The effect of finite momentum resolution on our model for the magnon dispersion is shown in Fig.~\ref{Qres}. 
Whilst we have not calculated it explicitly at the Os L$_3$ edge, we expect the momentum resolution of the spectrometer to be of the same order as that determined for the Ir L$_3$ edge.\cite{pincini2017, donnerer_phd}
Using the experimental value of the spin wave gap, and taking $\Delta Q=0.17\,\mathrm{\AA}^{-1}$, then we find that $\Gamma=\text{0.5(3)~meV}$ (Fig.~\ref{Qres}c). 
Notably our simulations reveal that the experimental data can be reasonably described by an \emph{isotropic} Hamiltonian ($\Gamma=0$) if $\Delta Q = \text{0.2 \AA}^{-1}$ (Fig.~\ref{Qres}b). 
Since anisotropy is partly governed by the spin-orbit interaction, then this implies that its effect upon the magnetic ground state and excitations in \naoso\ is lower than previously believed.
 
\section{Choice of incident energy}
In Fig.~\ref{XAS}, we compare the x-ray absorption (XAS) and RIXS signals at 300~K as a function of incident energy. The maximum in the RIXS signal (for fixed energy loss of 1~eV) occurs around 3~eV below the maximum of the white line; this mirrors the \emph{elastic} resonant magnetic scattering signal.\cite{calder2012} Because the main focus of our study was the low-energy spin excitations, we fixed the incident energy to 10.877~keV.
\\
\setcounter{figure}{0}
\begin{figure}[h!]
\includegraphics{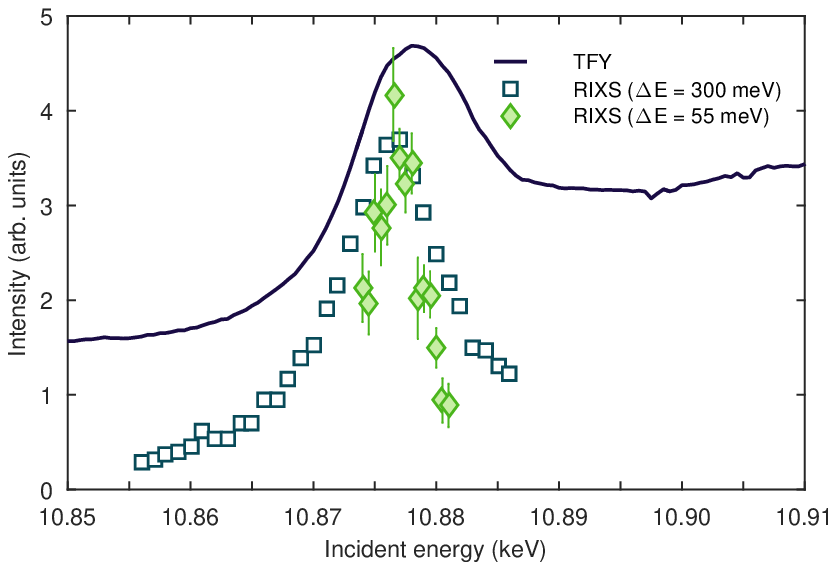}
\caption{Comparison of x-ray absorption (XAS) and RIXS signals at 300~K as a function of incident photon energy. All datasets have been scaled vertically to fit on the same axes. Solid line: XAS collected in total fluorescence yield (TFY) mode. Open squares: RIXS intensity using a low-resolution setup $(\Delta E = \text{300~meV})$. Filled diamonds: RIXS intensity using a high-resolution setup $(\Delta E = \text{55~meV})$. Both sets of RIXS data were collected for a fixed energy loss of 1~eV; this corresponds to the maximum of peak $\bm{\alpha}$.}
\label{XAS}
\end{figure} 

\section{Details of one- and two-magnon scattering cross-sections}
\label{appendix_mag}
Here we provide further details about the calculations performed in Section \ref{twomagnon_section}.

\subsection{One-magnon scattering cross-section}
\setcounter{figure}{0}
Within linear spin wave theory (LSWT), it is assumed that spin waves occur as a result of fluctuations of the magnetic moment transverse to the ordered spin direction. The spin raising and lowering operators are rewritten in terms of boson annihilation (creation) operators $\Alad{(\dag)}{i}$ and $\Blad{(\dag)}{i}$, using a Holstein-Primakoff transformation:

\arraycolsep=8pt
\begin{align}
\begin{array}{ll}
\mathbf{S}_i^+ = \sqrt{2S \left(\text{1}-\frac{\Alad{\dag}{i}\Alad{}{i}}{2S}\right)}\Alad{}{i} & \mathbf{S}_j^+ = \Blad{\dag}{j}\sqrt{2S \left(1-\frac{\Blad{\dag}{i}\Blad{}{i}}{2S}\right)} \\
\mathbf{S}_i^- = \Alad{\dag}{i}\sqrt{2S \left(1-\frac{\Alad{\dag}{i}\Alad{}{i}}{2S}\right)} & \mathbf{S}_j^- = \sqrt{2S \left(1-\frac{\Blad{\dag}{j}\Blad{}{j}}{2S}\right)}\Blad{}{j} \\
\mathbf{S}_i^z = S - \Alad{\dag}{i}\Alad{}{i} & \mathbf{S}_j^z = -S + \Blad{\dag}{i}\Blad{}{i},
\end{array}
\end{align}
where $\Alad{}{i}$ and $\Blad{}{i}$ operate on different magnetic sublattices.
In the Holstein-Primakoff approximation, the square root within the spin operators is formally expanded as a Taylor series into powers of $(\Alad{\dag}{i}\Alad{}{i}/\text{2}S)$ and $(\Blad{\dag}{i}\Blad{}{i}/\text{2}S)$. 

Neglecting magnon-magnon interactions, and taking only the leading order terms of the expansion, one obtains the approximate form of the spin operators within LSWT:
\arraycolsep=8pt
\begin{align}
\begin{array}{ll}
\mathbf{S}_i^+ \simeq \sqrt{\text{2}S}\,\Alad{}{i} & \mathbf{S}_j^+ \simeq \sqrt{\text{2}S}\,\Blad{\dag}{j} \\
\mathbf{S}_i^- \simeq \sqrt{\text{2}S}\,\Alad{\dag}{i} & \mathbf{S}_j^- \simeq \sqrt{\text{2}S}\,\Blad{}{j} \\
\mathbf{S}_i^z = S-\Alad{\dag}{i}\Alad{}{i} & \mathbf{S}_j^z = -S+\Blad{\dag}{j}\Blad{}{j}. \\
\end{array}
\label{LSWT_spinops}
\end{align}

Substituting these expressions for the spin operators into the magnetic Hamiltonian (Equation~\ref{ham_SM}), and then diagonalizing through a Bogoliubov transformation, one can determine the magnon dispersion relation and intensities.

The magnon dispersion relation $\omega(\mathbf{Q})$ is given by:        
\begin{align} \label{omega_SM}
\omega_{\mathbf{Q}} &= 2S\sqrt{A_{\mathbf{Q}}^2 - B_{\mathbf{Q}}^2} \nonumber \\
A_{\mathbf{Q}} &= 2J_1+J_2+3\Gamma \nonumber \\
B_{\mathbf{Q}} &= J_1\cos{\pi(h\!+\!l)} + J_1\cos{\pi(h\!-\!l)} +  J_2\cos{\pi k}.
\end{align}

For a $d^3$ system like NaOsO$_{\text{3}}$, one would expect $S=3/2$. In common with the analysis of the dispersion at 300~K presented in Ref.~\onlinecite{calder2017_naoso3}, we set $J_1=J_2=\text{14~meV}$ and $\Gamma=\text{1.4~meV}$.
\\

The scattering intensity of the single magnon mode depends on the Bogoliubov operators $u_{\mathbf{Q}}$ and $v_{\mathbf{Q}}$. These were used to diagonalize the Hamiltonian, and are defined as  $u_{\mathbf{Q}}=\cosh{\theta_{\mathbf{Q}}}$ and
$v_{\mathbf{Q}}=\sinh{\theta_{\mathbf{Q}}}$, with $\tanh{2\theta_{\mathbf{Q}}}=B_{\mathbf{Q}}/A_{\mathbf{Q}}$.
The corresponding transverse spin-spin correlations $S^{xx}\left(\mathbf{Q},\omega\right) = S^{yy}\left(\mathbf{Q},\omega\right)$ are given by:

\begin{align}
S^{xx} (\mathbf{Q},\omega) &\propto \frac{S-\Delta S^z}{2}\left[n(\omega_{\mathbf{Q}})+1\right](u_{\mathbf{Q}}-v_{\mathbf{Q}})^2 \nonumber \\
&\times\delta\!\left(\hbar\omega - \hbar\omega_{\mathbf{Q}}\right),
\end{align}
where $n(\omega_{\mathbf{Q}})$ is the Bose factor, and $\Delta S^z = S - \langle S^z \rangle$ is the spin reduction due to zero-point fluctuations.
Consequently to order $\Delta S^z$ the single magnon partial differential cross-section is given by:
\begin{widetext}
\begin{align}
\left(\frac{\text{d}^2\sigma}{\text{d}\Omega\text{d}E'}\right)^{\!\!(xx)} \propto \frac{S-\Delta S^z}{2} (u_{\mathbf{Q}}-v_{\mathbf{Q}})^2 \left\lbrace \left[n(\omega_{\mathbf{Q}})+1\right] \delta\!\left(\hbar\omega - \hbar\omega_{\mathbf{Q}}\right) + n(\omega_{\mathbf{Q}})\,\delta\!\left(\hbar\omega + \hbar\omega_{\mathbf{Q}}\right)\right\rbrace.
\end{align}
\end{widetext}
The total intensity of the transverse correlations integrated over energy and the entire Brillouin zone is given by $S^{xx}(\mathbf{Q},\omega) + S^{yy}(\mathbf{Q},\omega) = (S-\Delta S^z)(2\Delta S^z+1)$. This fact shall be used later to correctly normalize the relative contributions to the inelastic spectra.

\subsection{Two magnon cross-section}
The longitudinal component of the scattering cross-section $S^{zz}(\mathbf{Q},\omega)$ has two contributions: an elastic part ($\hbar\omega=\text{0}$), and an inelastic part ($\hbar\omega\neq \text{0}$). These shall be outlined in turn.

\subsubsection{Elastic component}
The elastic contribution gives the intensity of the magnetic Bragg reflection associated with the time-independent (i.e.~non-fluctuating) part of the spin operator $\mathbf{S}^z$.
This component of $S^{zz}(\mathbf{Q},\omega)$ shall be neglected for now since we are interested in the inelastic processes, but is included for completeness:
\begin{align}
\left(\frac{\text{d}^2\sigma}{\text{d}\Omega\text{d}E'}\right)^{\!\!(zz)}_{\hbar\omega=0} &= \quad \delta({\hbar\omega})\frac{\left(2\pi\right)^3}{V_m} \left(S-\Delta S^z \right)^2 \nonumber \\ &\times \quad \sum_{\bm{\tau}}\,\delta\!\left(\mathbf{Q} - \left(\bm{\tau} - \bm{\tau}_{\text{AFM}} \right) \right),
\end{align}
where $V_m$ is the volume of the magnetic unit cell, the sum over $\mathbf{\tau}$ extends over all structural Bragg reflections and $\bm{\tau}_{\text{AFM}}$ is the wavevector of any magnetic Bragg reflection. Note that the magnetic propagation vector $\mathbf{q}=0$ for \naoso. When integrated over energy and a single Brillouin zone, $S^{zz}\left(\mathbf{Q},\omega\right)_{\text{el}}=\left(S-\Delta S\right)^2$. 

\subsubsection{Inelastic contribution}
The inelastic contribution $S^{zz}\left(\mathbf{Q},\omega\right)_{\hbar\omega\neq 0}$ arises due to two-magnon scattering. 
There are three different types of two-magnon processes: two-magnon creation, two-magnon annihilation (both $\Delta S=2$), and mixed magnon creation-annihilation $(\Delta S=0)$. The partial differential two-magnon cross section is equal to the sum of these discrete processes, which are given below.
At low temperatures only the two magnon creation term (Eqn.~\ref{twomag_cc}) contributes significantly to the scattering cross-section, since $\bose\approx 0$. At elevated temperatures however the other components also become important.
The above expressions are valid if magnon-magnon interactions are neglected.
When integrated over energy and a single Brillouin zone, $S^{zz}(\mathbf{Q},\omega)_{\hbar\omega\neq 0} = \Delta S^z(1-\Delta S^z) + \langle uv\rangle^2$, where $\langle uv\rangle^2 = \sum_{\mathbf{q}} u_{\mathbf{q}}v_{\mathbf{q}} / N$.

\begin{widetext}
\begin{align}
\left(\frac{\text{d}^2\sigma}{\text{d}\Omega\text{d}E'}\right)^{\!\!(zz)}_{\hbar\omega\neq 0} &= \left(\frac{\text{d}^2\sigma}{\text{d}\Omega\text{d}E'}\right)^{\!\!(zz)}_{\!\!(aa)} +
\left(\frac{\text{d}^2\sigma}{\text{d}\Omega\text{d}E'}\right)^{\!\!(zz)}_{\!\!(cc)} +
 \left(\frac{\text{d}^2\sigma}{\text{d}\Omega\text{d}E'}\right)^{\!\!(zz)}_{\!\!(ca)} \label{twomag_weight}  \\
 \left(\frac{\text{d}^2\sigma}{\text{d}\Omega\text{d}E'}\right)^{\!\!(zz)}_{\!\!(aa)} &\propto
 \frac{2}{N}\sum_{\mathbf{q}} \frac{1}{2}\left(\bogo{u}\bogoqq{v} - \bogoqq{u}\bogo{v} \right)^2 \times \bose\left[\boseqq + 1\right] \delta\!\left(\hbar\omega + \hbar\omega_{\mathbf{q}}+\hbar\omega_{\mathbf{q}+\mathbf{Q}}\right) \\
 \left(\frac{\text{d}^2\sigma}{\text{d}\Omega\text{d}E'}\right)^{\!\!(zz)}_{\!\!(cc)} &\propto
 \frac{2}{N}\sum_{\mathbf{q}} \frac{1}{2}\left(\bogo{u}\bogoqq{v} - \bogoqq{u}\bogo{v} \right)^2 \times \left[\bose + 1\right]\left[\boseqq + 1\right] \delta\!\left(\hbar\omega - \hbar\omega_{\mathbf{q}}-\hbar\omega_{\mathbf{q}+\mathbf{Q}}\right) \label{twomag_cc}\\
 \left(\frac{\text{d}^2\sigma}{\text{d}\Omega\text{d}E'}\right)^{\!\!(zz)}_{\!\!(ca)} &\propto
 \frac{2}{N}\sum_{\mathbf{q}} \left(\bogo{u}\bogoqq{u} - \bogo{v}\bogoqq{v} \right)^2 \times \bose\left[\boseqq + 1\right] \delta\!\left(\hbar\omega + \hbar\omega_{\mathbf{q}}-\hbar\omega_{\mathbf{q}+\mathbf{Q}}\right).
\end{align}
\end{widetext}
%

\end{document}